\documentclass[manuscript]{emulateapj}

\shorttitle{The Herschel ATLAS}
\shortauthors{Eales et al.}

\begin{document}


\title{The Herschel ATLAS}


\author{S. Eales\altaffilmark{1},
L. Dunne\altaffilmark{2},
D. Clements\altaffilmark{3},
A. Cooray\altaffilmark{4},
G. De Zotti\altaffilmark{5},
S. Dye\altaffilmark{1},
R. Ivison\altaffilmark{6},
M. Jarvis\altaffilmark{7},
G. Lagache\altaffilmark{8},
S. Maddox\altaffilmark{2},
M. Negrello\altaffilmark{9},
S. Serjeant\altaffilmark{9},
M.A. Thompson\altaffilmark{7},
E. Van Kampen\altaffilmark{10},
A. Amblard\altaffilmark{4},
P. Andreani\altaffilmark{10},
M. Baes\altaffilmark{11},
A. Beelen\altaffilmark{8},
G.J. Bendo\altaffilmark{3},
D. Benford\altaffilmark{12},
F. Bertoldi\altaffilmark{40},
J. Bock\altaffilmark{47},
D. Bonfield\altaffilmark{7},
A. Boselli\altaffilmark{14},
C. Bridge\altaffilmark{13},
V. Buat\altaffilmark{14},
D. Burgarella\altaffilmark{14},
R. Carlberg\altaffilmark{48},
A. Cava\altaffilmark{15},
P. Chanial\altaffilmark{3},
S. Charlot\altaffilmark{16},
N. Christopher\altaffilmark{45},
P. Coles\altaffilmark{1},
L. Cortese\altaffilmark{1},
A. Dariush\altaffilmark{1},
E. da Cunha\altaffilmark{17},
G. Dalton\altaffilmark{18},
L. Danese\altaffilmark{19},
H. Dannerbauer\altaffilmark{20},
S. Driver\altaffilmark{46},
J. Dunlop\altaffilmark{31},
L. Fan\altaffilmark{19},
D. Farrah\altaffilmark{21},
D. Frayer\altaffilmark{22},
C. Frenk\altaffilmark{23},
J. Geach\altaffilmark{23},
J. Gardner\altaffilmark{12},
H. Gomez\altaffilmark{1},
J. Gonz\'alez-Nuevo\altaffilmark{19},
E. Gonz\'alez-Solares\altaffilmark{24},
M. Griffin\altaffilmark{1},
M. Hardcastle\altaffilmark{7},
E. Hatziminaoglou\altaffilmark{10}, 
D. Herranz\altaffilmark{25},
D. Hughes\altaffilmark{26},
E. Ibar\altaffilmark{6},
Woong-Seob Jeong\altaffilmark{50},
C. Lacey\altaffilmark{23},
A. Lapi\altaffilmark{51},
M. Lee\altaffilmark{27},
L. Leeuw\altaffilmark{28},
J. Liske\altaffilmark{10},
M. L\'opez-Caniego\altaffilmark{29},
T. M\"uller\altaffilmark{30},
K. Nandra\altaffilmark{3},
P. Panuzzo\altaffilmark{39},
A. Papageorgiou\altaffilmark{1},
G. Patanchon\altaffilmark{49},
J. Peacock\altaffilmark{31},
C. Pearson\altaffilmark{32},
S. Phillipps\altaffilmark{33},
M. Pohlen\altaffilmark{1},
C. Popescu\altaffilmark{34},
S. Rawlings\altaffilmark{45},
E. Rigby\altaffilmark{2},
M. Rigopoulou\altaffilmark{18},
G. Rodighiero\altaffilmark{42},
A. Sansom\altaffilmark{34},
B. Schulz\altaffilmark{13},
D. Scott\altaffilmark{35},
D.J.B. Smith\altaffilmark{2},
B. Sibthorpe\altaffilmark{6},
I. Smail\altaffilmark{23},
J. Stevens\altaffilmark{7},
W. Sutherland\altaffilmark{36},
T. Takeuchi\altaffilmark{37},
J. Tedds\altaffilmark{38},
P. Temi\altaffilmark{28},
R. Tuffs\altaffilmark{41},
M. Trichas\altaffilmark{3},
M. Vaccari\altaffilmark{42},
I. Valtchanov\altaffilmark{43},
P. van der Werf\altaffilmark{44},
A. Verma\altaffilmark{45},
J. Vieria\altaffilmark{13},
C. Vlahakis\altaffilmark{44} \&
Glenn J. White\altaffilmark{9,32}
}


\altaffiltext{1}{School of Physics and Astronomy, Cardiff University,
Queens Buildings, The Parade, Cardiff CF24 3AA, UK}
\altaffiltext{2}{School of Physics and Astronomy, University of Nottingham, Nottingham, NG7 2RD, UK} 
\altaffiltext{3}{Physics Department, Imperial College London, Prince Consort Road, London, SW7 2AZ, UK}
\altaffiltext{4}{Center for Cosmology, University of California, Irvine, CA 92697, USA}
\altaffiltext{5}{INAF-Osservatorio Astronomico di Padova, I-35122 Padova, and SISSA, I-34014 Trieste, Italy}
\altaffiltext{6}{UK Astronomy Technology Centre, Royal Observatory, Blackford Hill, Edinburgh EH9 3HJ, UK}
\altaffiltext{7}{Centre for Astrophysics Research, STRI, University of Hertfordshire, Hatfield, AL10 9AB, UK}
\altaffiltext{8}{Institut d'Astrophysique Spatiale (IAS), Bâtiment 121, F-91405 Orsay, France; and Université Paris-Sud 11 and CNRS (UMR 8617), France}
\altaffiltext{9}{Dept. of Physics and Astronomy, The Open University, Milton Keynes, MK7 6AA, UK}
\altaffiltext{10}{European Southern Observatory, Karl-Schwarzschild-Str. 2, 85748 Garching bei M{\"u}nchen, Germany}
\altaffiltext{11}{Sterrenkundig Observatorium, Universiteit Gent, Krijgslaan 281 S9, B-9000 Gent, Belgium}
\altaffiltext{12}{Observational Cosmology Laboratory, Code 665, NASA's Goddard Space Flight Center, Greenbelt MD 20771, USA}
\altaffiltext{13}{California Institute of Technology, 1200 East California Blvd., Pasadena, CA 91125, USA}
\altaffiltext{14}{Laboratoire d'Astrophysique de Marseille, Observatoire Astronomique Marseille Provence, Aix-Marseille Université, CNRS, Franc}
\altaffiltext{15}{Instituto de Astrofísica de Canarias, Calle Vía Láctea, E38205, La Laguna, Esp}
\altaffiltext{16}{Institut d'Astrophysique de Paris, CNRS UMR7095 UPMC, 98 bis boulevard Arago, F-75014 Paris, France}
\altaffiltext{17}{Department of Physics, University of Crete, 71003 Heraklion, Greece}
\altaffiltext{18}{Astrophysics, University of Oxford, Oxford OX1 3RH and The Rutherford Appleton Laboratory, Chilton, Didcot, OX11 0QX}
\altaffiltext{19}{SISSA, Via Beirut 2-4, I-34014 Trieste, Italy}
\altaffiltext{20}{Max Planck Institute for Astronomy, Königstuhl 17, D-69117 Heidelberg, German}
\altaffiltext{21}{Astronomy Centre, University of Sussex, Brighton, UK}
\altaffiltext{22}{Infrared Processing and Analysis Center, California Institute of Technology 100-22, Pasadena, CA 91125, USA}
\altaffiltext{23}{Institute for Computational Cosmology, Physics Dept, Durham University, South Road, Durham DH1 3LE}
\altaffiltext{24}{Institute of Astronomy, University of Cambridge, Madingley Rd., Cambridge CB3 0HA, UK}
\altaffiltext{25}{Instituto de F\'isica de Cantabria (CSIC-UC), Santander, 39005, Spain}
\altaffiltext{26}{Instituto Nacional de Astrof\'isica, Optica y Electr\'onica (INAOE), Aptdo. Postal 51 y 216, Puebla, Mexico}
\altaffiltext{27}{Astronomy Program, Department of Physics and Astronomy, Seoul National University, Seoul 151-742, KOREA}
\altaffiltext{28}{Astrophysics Branch, NASA Ames Research Center, Mail Stop 245-6, Moffett Field, CA 94035, USA}
\altaffiltext{29}{Instituto de F\'isica de Cantabria (CSIC-UC), Santander, 39005, Spain}
\altaffiltext{30}{Max-Planck-Institut f\"ur extraterrestrische Physik, Giessenbachstrasse, 85748 Garching, Germany}
\altaffiltext{31}{Institute for Astronomy, University of Edinburgh, Royal Observatory, Edinburgh EH9 3HJ, UK}
\altaffiltext{32}{Rutherford Appleton Laboratory, Chilton, Didcot, Oxfordshire OX11 0QX, UK}
\altaffiltext{33}{Astrophysics Group, Department of Physics, University of Bristol, Tyndall Avenue, Bristol BS8 1TL}
\altaffiltext{34}{Jeremiah Horrocks Institute, University of Central Lancashire, Preston, PR1 2HE, UK}
\altaffiltext{35}{Department of Physics \& Astronomy, University of British Columbia, 6224 Agricultural Road, Vancouver BC, V6T1Z1, Canada}
\altaffiltext{36}{Astronomy Unit, Queen Mary University of London, Mile End Road, London E1 4NS, UK}
\altaffiltext{37}{Institute for Advanced Research, Nagoya University, Furo-cho, Chikusa-ku, Nagoya 464-8601, Japan}
\altaffiltext{38}{Department of Physics and Astronomy, University of Leicester, University Road, Leicester LE1 7RH, UK}
\altaffiltext{39}{CEA, Laboratoire AIM, Irfu/SAp, F-91191 Gif-sur-Yvette, France}
\altaffiltext{40}{Argelander Institute for Astronomy, Bonn University, Auf dem Huegel 71, D-53121 Bonn, Germany}
\altaffiltext{41}{Max Planck Institut f\"ur Kernphysik, Saupfercheckweg 1, D-69117, Heidelberg, Germany}
\altaffiltext{42}{Department of Astronomy, University of Padova, Vicolo Osservatorio 3, I-35122, Padova, Italy}
\altaffiltext{43}{Herschel Science Centre, ESA, P.O. Box 78, 28691 Villanueva de la Cañada, Madrid, Span} 
\altaffiltext{44}{Leiden Observatory, Leiden University, P.O. Box 9513, NL - 2300 RA Leiden, The Netherlands}
\altaffiltext{45}{Oxford Astrophysics, Denys Wilkinson Building, University of Oxford, Keble Road, Oxford, OX1 3RH}
\altaffiltext{46}{School of Physics and Astronomy, University of St Andrews, St Andrews, KY16 9SS}
\altaffiltext{47}{Astrophysics and Space Science, Jet Propulsion Laboratory, Pasadena, CA, 91109, USA; and Department of Physics, Math and Astronomy, California Institute of Technology, Pasadena, CA, 91125, USA}
\altaffiltext{48}{Department of Astronomy and Astrophysics, 50 St. George Street,
Toronto, Ontario, M5S 3H4, Canada}
\altaffiltext{49}{Universit\'e Denis Diderot, Laboratoire Astro-Particules et Cosmologie, 10 rue Alice Domon et L\'eonie Duquet, 75205, Paris Cedex 13, France}
\altaffiltext{50}{Space Science Division, Korea Astronomy \& Space Science Institute, 61-1, Whaam-dong, Yuseung-gu, Deajeon, 305-348, Republic of Korea}
\altaffiltext{51}{Physics Dept., University Tor Vergata, Via della Ricerca Scientifica 1, I-00133 Roma, Italy }


\begin{abstract}
The Herschel ATLAS is the largest open-time key project that will be carried out on the
Herschel Space Observatory. It will survey 510 square degrees of the extragalactic
sky, four times larger than all the other Herschel surveys combined, in five far-infrared
and submillimetre bands.
We describe the survey, the complementary multi-wavelength datasets that will
be combined with the Herschel data, and the six major science programmes we are undertaking.
Using new models based on a previous submillimetre survey of galaxies, we present predictions of the properties
of the ATLAS sources in other wavebands.

\end{abstract}

\section{Introduction}

Approximately half the energy emitted since the big bang by all the objects in the
Universe has been absorbed by dust and then reradiated between 60 and 500 $\mu$m \citep{dwek98,fix98,driver1},
a wavelength range in which the Universe is still
largely unexplored (Fig. 1). On the short-wavelength side of this waveband, 
the whole sky was surveyed at 60 and 100 $\mu$m by IRAS in the 1980s.
However, almost all of the tens of thousands of
galaxies detected by IRAS were spirals and starbursts in the
nearby Universe ($\rm z < 0.1$), and IRAS revealed
little about the dust in other galaxy populations, especially early-type galaxies
\citep{breg98}.
Even in the late-type galaxies, only the small fraction of the the dust warm enough to radiate
significantly in the far-infrared was detected by IRAS. Devereux and Young (1990), for example, 
showed that the
gas-to-dust ratio estimated from IRAS measurements alone was $\simeq$10 times greater than the standard
Galactic value, implying that $\simeq$90\% of the
dust in galaxies was effectively missed by IRAS. ISO and Spitzer, with their
long wavelength (170 $\mu$m) band, suffered less from this problem but will still have effectively missed any dust
with $\rm T < 15 K$ \cite{ben2003}.
Notwithstanding the successes of IRAS, ISO and Spitzer, most of the waveband from 60 to 500
$\mu$m 
is still virtually
{\it terra incognita}, and the only survey of a large area of the extragalactic sky
at a wavelength beyond 200 $\mu$m is the one recently carried
out by the Herschel pathfinder experiment, the Balloon Large
Area Survey Telescope (BLAST), which covers
$\simeq$ 20 deg$^2$ \citep{devlin}. 

The lack of a survey covering a large area of sky in the submillimetre waveband
(in this paper defined as $\rm 100\ \mu m < \lambda < 1\ mm$) has left us in some ways
knowing more about dust in the distant, early Universe than in the Universe
today.
The surveys that have been carried out
in the submillimetre waveband with ground-based telescopes---at 450 and
850 $\mu$m with the
SCUBA camera on the James Clerk Maxwell Telescope \cite{h98,eales99,cop2006},
at 1.2 mm with MAMBO on the IRAM 30-m telescope \cite{greve2004,bert2007,greve2008} and at 1.1 mm
with AZTEC on the James Clerk Maxwell Telescope \cite{per08,scott08}---have been of 
very small areas of sky, covering $\rm \sim 1\ deg^2$ of sky in total. Because of the
unusual
submillimetre `K-correction'\footnote{Beyond a redshift of $\simeq1$, as the redshift increases, the typical
spectral energy distribution of a dusty galaxy means that the effect of
increasing luminosity-distance on the brightness of the galaxy is compensated for by the increasing
rest-frame luminosity of the galaxy. The consequence is that the galaxy's flux density is approximately
independent of redshift.},
these surveys have mostly detected sources at very
high redshifts ($\rm z \succeq 1$). They have
led to the important discovery
that there is a population of luminous dust-enshrouded galaxies in the early Universe
\cite{sib,h98}, which many authors have suggested
are the
ancestors of ellipticals today
\cite{scott02,dunne2003}, but they have told us relatively little about 
the evolution of the Universe since $\rm z \simeq 1$ (the last 8 billion years)
and almost nothing at all about dusty galaxies in the nearby Universe.

\begin{figure}
\figurenum{1}
\plotone{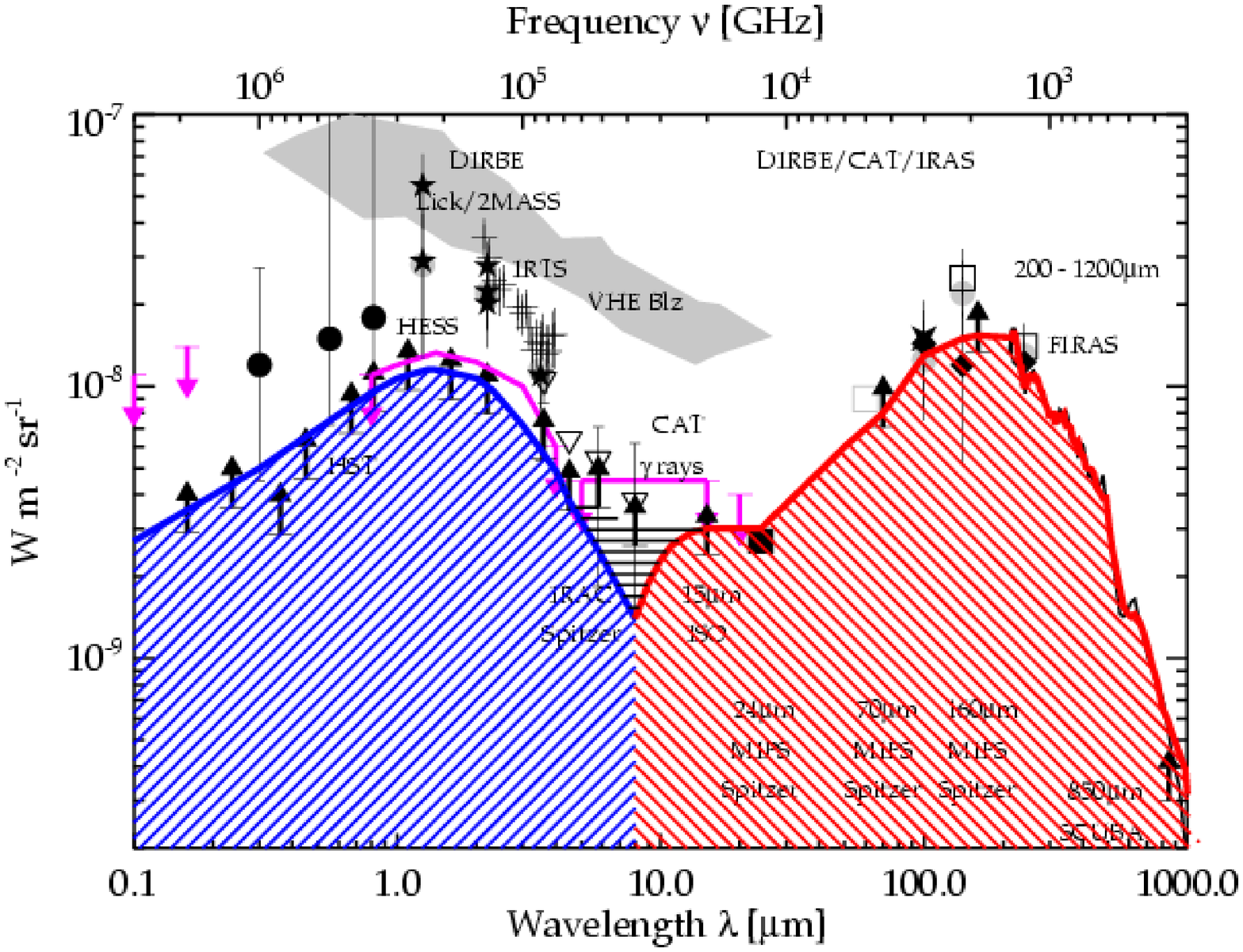}
\caption{The extraglactic background radiation as a function of wavelength \citep{dole}}
\end{figure}

Two basic things one would like to know about the nearby Universe
are the submillimetre luminosity function and the dust-mass function(the space-density
of galaxies as a function of dust mass). These functions are important for many reasons,
including tests of 
semi-analytical models of galaxy formation \cite{cole,baugh} and, by comparison with the same functions at
high redshift,  accurate measurements
of the
amount of evolution that is occurring in the submillimetre waveband \citep{dunne2003}.
Unfortunately, our knowledge of these functions is still extremely poor because
of the limitations in areal coverage and sensitivity of previous submillimetre
telescopes.
Until recently the only estimates of the local submillimetre luminosity
function were one based on 55 galaxies detected in an ISO
170-$\mu$m survey \citep{tak2006} and ones based on SCUBA 850-$\mu$m observations of 
$\simeq$200 galaxies selected in other wavebands \citep{dunne2000,cat2005}.
There are now direct estimates from the BLAST results of the local luminosity function at 250, 350 and
500 $\mu$m \citep{eales09}, but their accuracy is limited by the small
number of low-redshift sources detected in the BLAST survey: $\rm \simeq 30$ at $\rm z < 0.2$. 

The lack of a large-area survey capable of measuring the dust content and dust-obscured star formation 
in large numbers of galaxies in the local Universe has been especially galling for
submillimetre astronomers in the light of the success of their colleagues working in other
wavebands. The Sloan Digital Sky Survey and the 2dF Galaxy Redshift Survey have led
to a revolution in our understanding of the distribution of galaxies in the local
Universe, and the relationships between their present star-formation rate, star-formation
history, stellar mass, morphology and environment \cite{lew02,kauff03a,kauff03b,alan04,balogh04}. 
However, all the studies that have used these impressive datasets to
investigate the physics and ecology of the 
galaxy population have been forced to ignore 
the dust phase of the interstellar medium and star formation that is heavily obscured by dust,
because the IRAS survey was only sensitive enough
to detect a small percentage of the galaxies in the redshift
surveys (1.8 per cent of the SDSS galaxies - Obric et al. 2006), and 
it missed 90\% 
of the dust in the detected galaxies because of its insensitivity to cold dust \citep{dy90}.

The launch of the Herschel Space Observatory, which occurred on May 14th 2009,
has the potential to dramatically increase our knowledge
of dust and dust-obscured star formation, especially in the nearby Universe. Herschel
has two main cameras: SPIRE, which will be able to
image the sky simultaneously at 250,
350 and 500 $\mu$m \cite{matt}, and PACS, which will be able to image the sky in two bands simultaneously,
either 70 and 170 $\mu$m or 110 and 170 $\mu$m \cite{pog}. Herschel will have much better angular resolution
($\simeq$18 arcsec at 250 $\mu$m) and sensitivity than previous observatories, and the spectral coverage
of SPIRE will make it possible to carry out the first large-area surveys in this virtually
unexplored part of the electromagnetic spectrum. The SPIRE bands will also make it possible to
detect the cold dust that was missed by earlier observatories. Herschel is also better suited
for investigating the local Universe than the submillimetre surveys that will soon be carried out
from the ground, in particular the SCUBA-2 and LABOCA surveys, because these will operate mostly
at 850 $\mu$m, where low-redshift galaxies are intrinsically faint, whereas the Herschel bands span the peak of the
typical spectral energy distribution of a galaxy in the nearby Universe. 

In this paper, we describe the largest project that will be carried out 
with the Herschel Space Observatory in `Open Time', the
time available for competition within the international
astronomical community\footnote{This forms
about 2/3 of the total observing time on Herschel, with the
remaining 1/3 being time reserved for the
teams that built the instruments---`Guaranteed Time'.}. The Herschel 
Astrohysical Terahertz Large Area Survey (the Herschel ATLAS or H-ATLAS)
will be a survey of
510 deg$^2$ of sky in five
photometric bands. This is $\simeq$8 times larger than the coverage
of the next largest (in area) Herschel extragalactic survey, HERMES (Oliver et al. 2009). 
The main scientific goal of the H-ATLAS is to provide measurements
of the dust masses and dust-obscured star formation for tens of thousands of nearby
galaxies, the far-IR/submillimetre equivalent to the SDSS photometric survey. However,
the H-ATLAS has many other science goals ranging from the investigation
of the point sources that will be detected by the Planck Surveyor
to a study of high-latitude galactic dust. 

The arrangement of this paper
is as follows. In Section 2 we describe the basic parameters of the survey.
In Section 3 we present predictions of the number and redshift distribution of
the sources that will be detected by the H-ATLAS.
In Section 4, 
we describe the complementary data that exists or will soon exist
for the H-ATLAS fields.
In Section 5 we describe the six main H-ATLAS science programmes.
In Section 6 we describe the detailed survey strategy, including issues
that will
be of interest to the general community,
such as our plans for the release of data products. We everywhere assume the
cosmological parameters for a `concordance universe': $\rm \Omega_M = 0.267$ and
$\rm \Omega_{\Lambda} = 0.762$. 

\begin{figure*}
\figurenum{2}
\plottwo{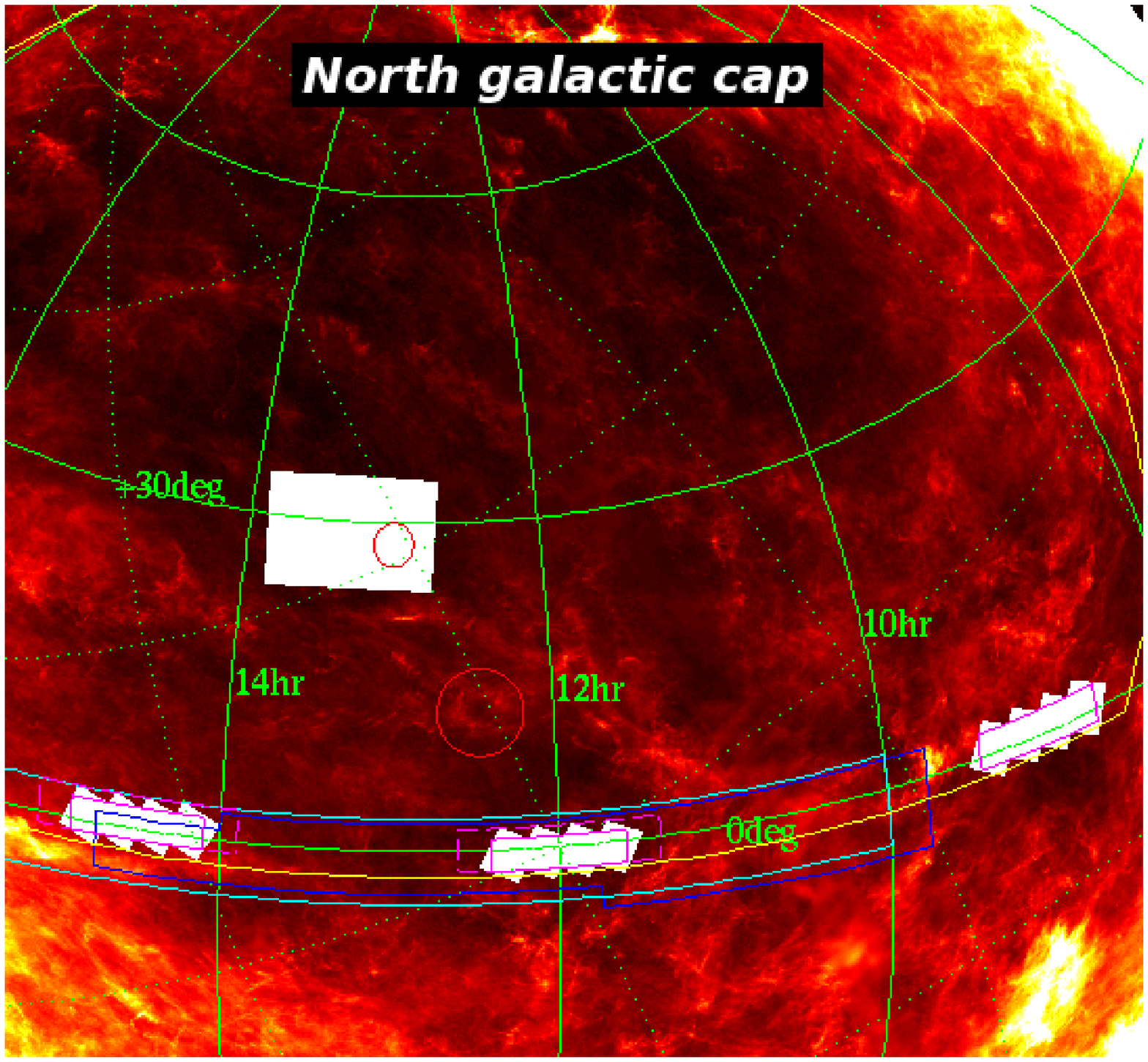}{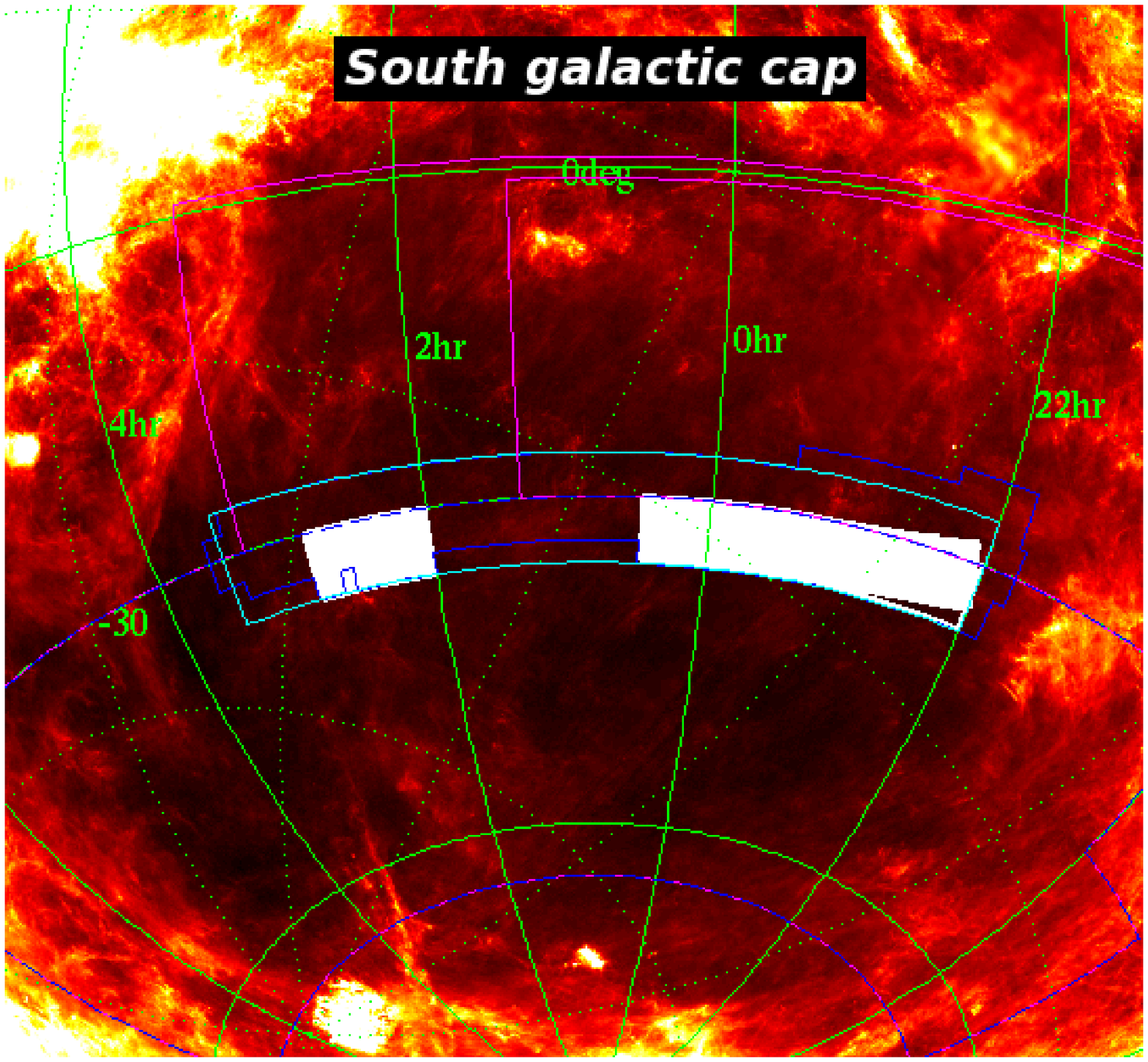}
\caption{The positions of the ATLAS field, shown as white blocks,
 superimposed on the IRAS 100 $\mu$m map
of the sky, which traces the distribution of galactic dust. Figure 2(a) shows
the northern galactic cap and Figure 2(b) shows the southern galactic cap.
The colour coding for the lines is as follow: solid green lines---RA and dec;
dotted green lines---ecliptic latitude and longitude; cyan---KIDS/VIKING area; yellow---SDSS area; blue---2dFGRS area;
magenta---area of the Dark Energy Survey; magenta/blue dashed---area covered by the South Pole Telescope.}
\end{figure*}

\section{The Basic Parameters of the Survey}

The H-ATLAS has been allocated 600 hours of time, making it the largest 
key project that will be carried out with Herschel in Open Time.
For all of our science goals (\S 5), the final sensitivity of the survey is
not critical and it is more important to survey the greatest possible
area of sky.
We have therefore chosen to
use the maximum possible scan rate for the telescope (60 arcsec sec$^{-1}$).
For our first science programme (\S 5.1), it is important to make observations with
PACS and SPIRE, and therefore we have chosen to use the Herschel observing mode
that allows simultaneous observations with the two cameras: Parallel Mode (PMode). 
Of the two possible combinations of photometric bands for PACS \cite{pog},
we have chosen to observe at 110 and 170 $\mu$m rather than at 70 and 170 $\mu$m mostly
on the grounds of sensitivity; the noise at 70 and 110 $\mu$m should be fairly similar but galaxies, 
even at low redshift, are generally brighter at the longer wavelength.
Although this combination will be worse for estimating the temperature
of the dust, our models suggest that we will still be able to obtain useful
measurements of the
temperature of the dust in low-redshift galaxies.

\begin{deluxetable}{cccc}
\tabletypesize{\scriptsize}
\tablecaption{H-ATLAS Fields}
\tablewidth{0pt}
\tablehead{
\colhead{Name} & \colhead{Centre} & \colhead{RA width$^a$} & \colhead{Dec width$^a$} \\
}
\startdata
NGP & 13 18 00, 29 00 00 & 15 & 10 \\
GAMA A & 09 00 00, 00 00 00 & 12 & 3 \\
GAMA B & 12 00 00, 00 00 00 & 12 & 3 \\
GAMA C & 14 30 00, 00 00 00 & 12 & 3 \\
SGP A & 02 26 48, -33 00 00& 11 & 6 \\
SGP B & 23 15 36, -32 54 00& 31 & 6 \\
\enddata
\tablenotetext{a}{The precise coverage of the fields and their orientation
on the sky will depend on exactly when they are observed. See \S 6 for
more details.}
\tablecomments{Reading from the left, the columns are:
the name of the field; the central position of the field;
the width of the field in degrees in RA; the width of the field in
degrees in
declination.}
\end{deluxetable}

With an eye on the legacy value of the H-ATLAS, we have chosen to observe fields in
the northern and southern hemispheres and on
the celestial equator. Other than that,
we have chosen our fields to maximise the amount of complementary data and to minimize the
amount of confusing emission from dust in the Galaxy, this last determined from the
IRAS 100 $\mu$m maps. The fields, which are shown in Figure 2 and listed
in Table 1, are:

\begin{itemize}

\item One field close to the northern galactic pole with an area of
150 deg$^2$ (henceforth the NGP field)

\item Three fields, each of 36 deg$^2$ in area, coinciding with the fields being surveyed
in the Galaxy And Mass Assembly redshift survey (Driver et al. 2009) (henceforth
the GAMA fields)

\item Two fields with a total area of 250 deg$^2$ close to the south galactic pole (henceforth the
SGP fields)

\end{itemize}

The total survey covers 510 deg$^2$. 
The angular resolution (full-width half maximum) of the observations will be approximately
8, 12, 18, 25 and 36 arcsec at 70, 110, 250, 350 and 500 $\mu$m, respectively.
The 5$\sigma$
sensitivities that should be reached in the
five bands are 67 mJy at 110 $\mu$m, 94 mJy at 170
$\mu$m, 45 mJy at 250 $\mu$m, 62 mJy at 350 $\mu$m and
53 mJy at 500 $\mu$m. These sensitivities have been
estimated using the current version of the Herschel observation planning package
HSpot. We have assumed that the sources
will be unresolved by the telescope beam, which will be true for most sources but will not be true for
the closest sources.
We have also neglected the effects of emission from dust in the Galaxy and of confusion from extragalactic
sources. The importance of these is still uncertain, but the importance of confusion should
be less than for the deeper Herschel surveys \citep{oliver}.

\begin{figure*}
\figurenum{3a}
\plotone{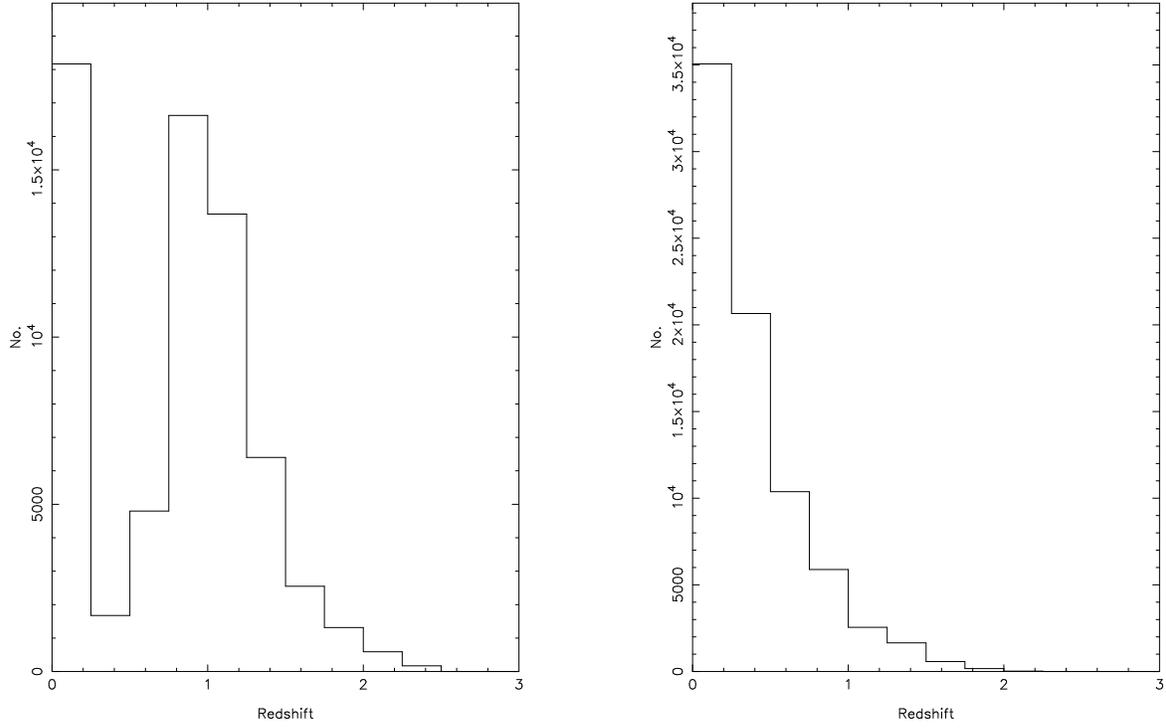}
\caption{On the left is the redshift distributions predicted at 110 $\mu$m by the models
of Lagache et al. (2003,2004); on the right is the redshift distribution predicted
at the same wavelength by the model described in the text.}
\end{figure*}

\begin{figure*}
\figurenum{3b}
\plotone{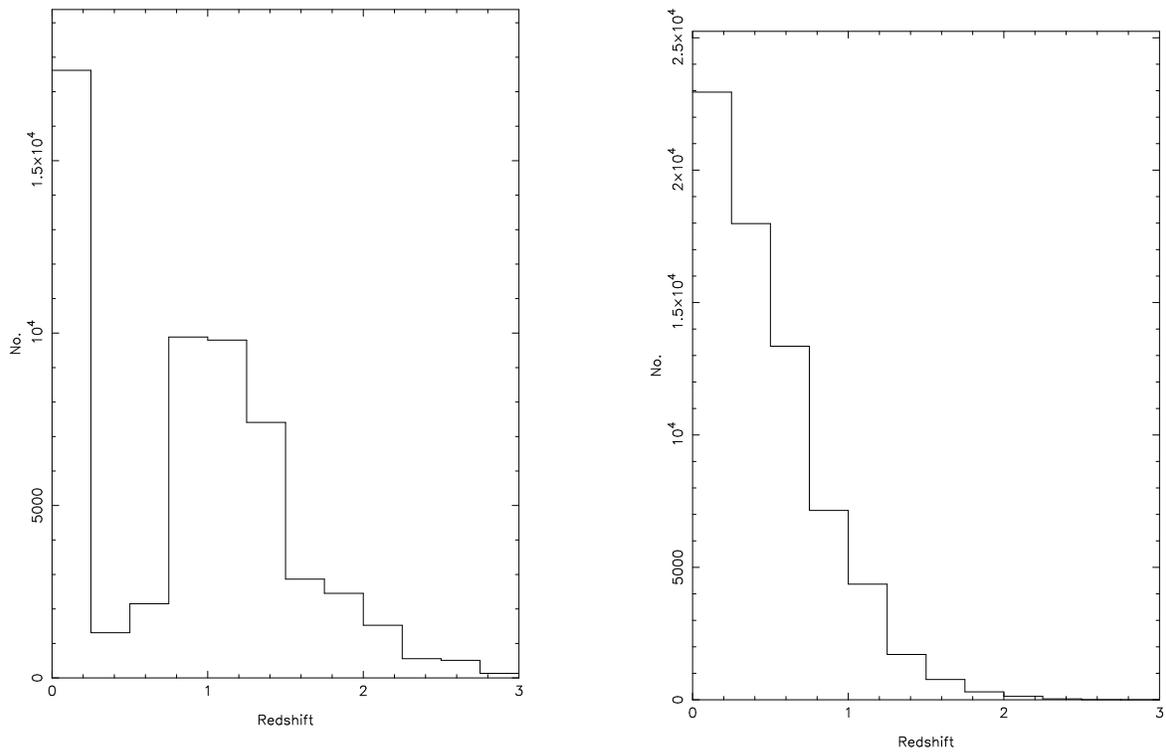}
\caption{Same as in (a) except at 170 $\mu$m.}
\end{figure*}

\begin{figure*}
\figurenum{3c}
\plotone{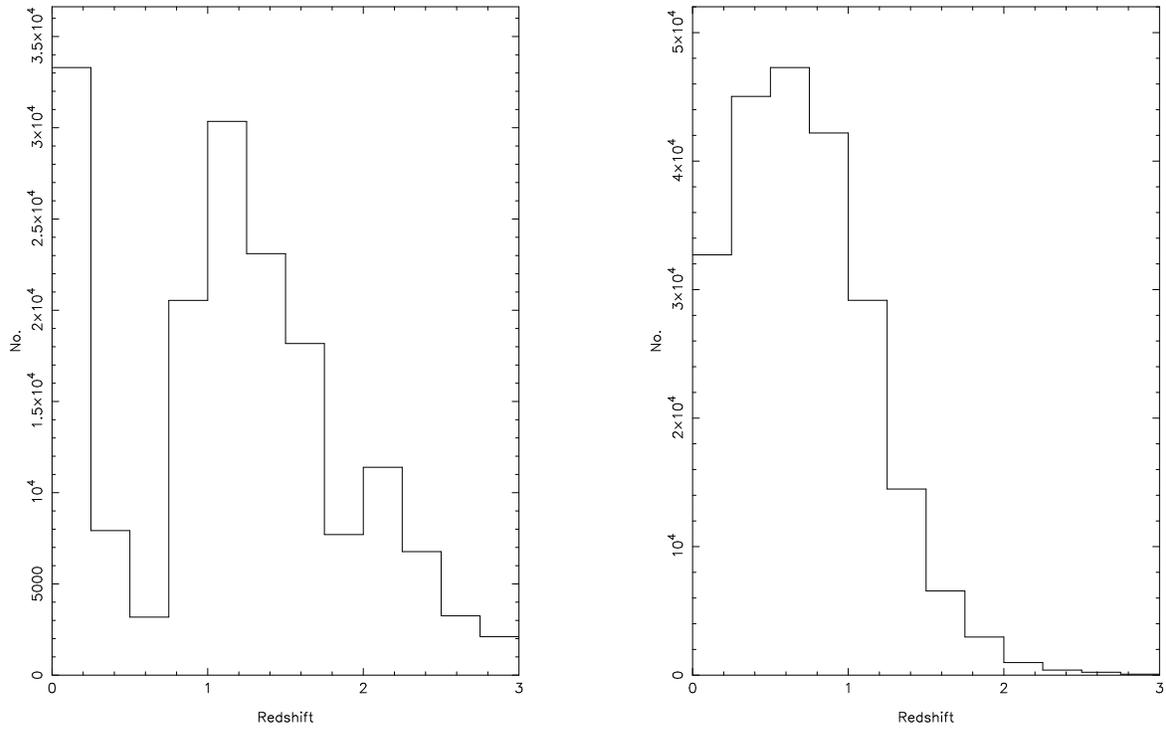}
\caption{Same as in (a) except at 250 $\mu$m.}
\end{figure*}

\begin{figure*}
\figurenum{3d}
\plotone{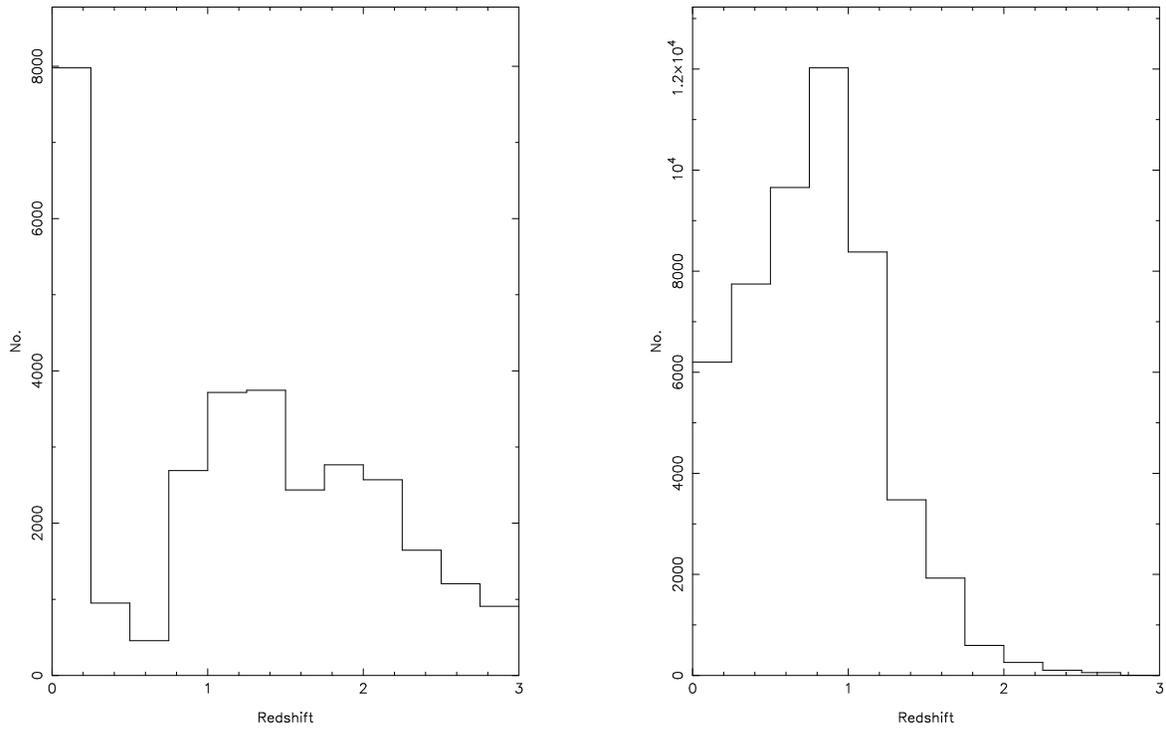}
\caption{Same as in (d) except at 350 $\mu$m.}
\end{figure*}

\begin{figure*}
\figurenum{3e}
\plotone{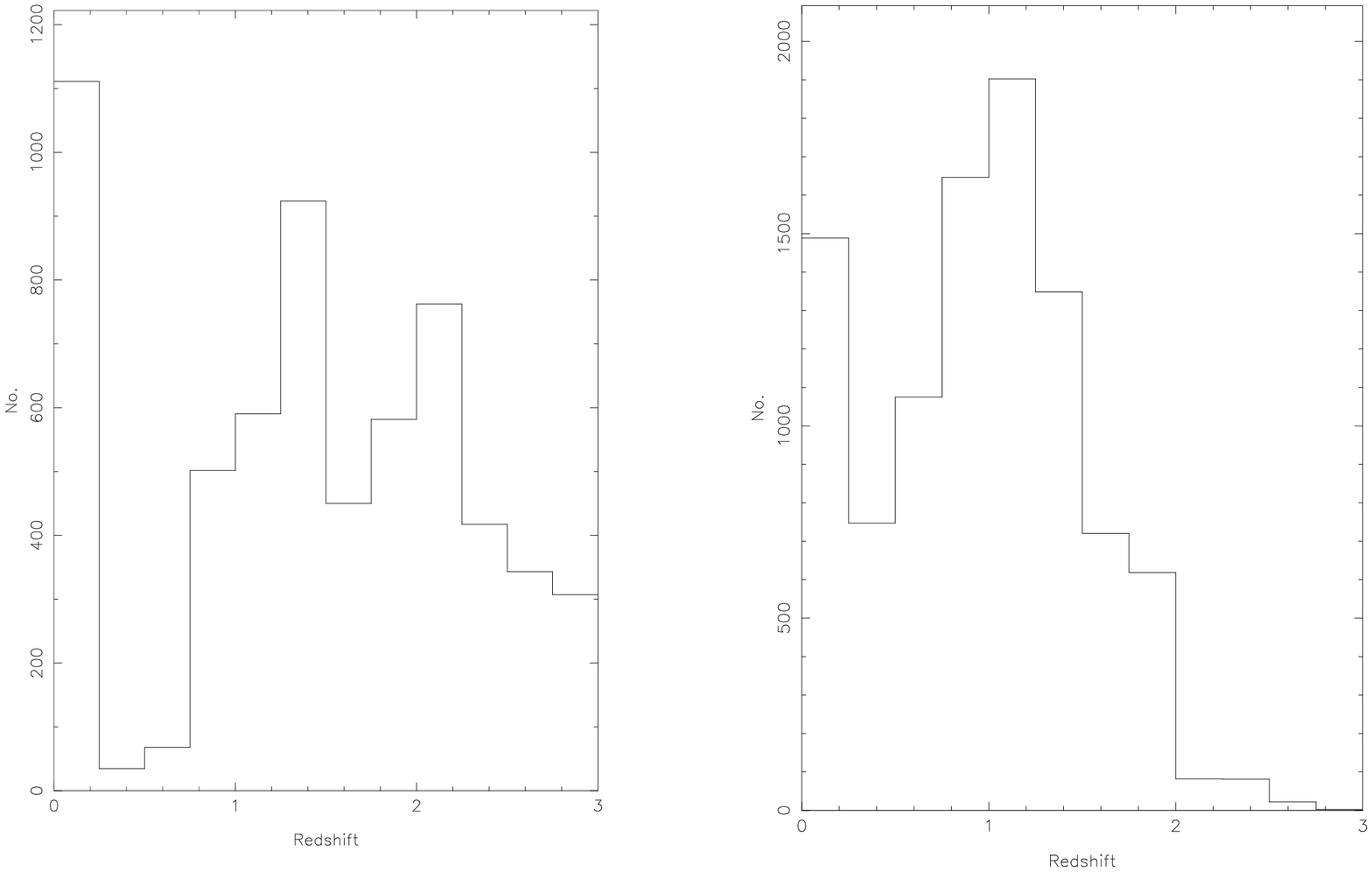}
\caption{Same as in (e) except at 500 $\mu$m.}
\end{figure*}

\section{Predictions}

The number of sources that will be found in any of the Herschel surveys is
uncertain because they cover regimes of flux density and wavelength that have never
been explored before. There is a particular problem in the wavelength range $\rm 200 <
\lambda < 500\ \mu m$ because no survey, until the very recent survey with 
the balloon-borne telescope BLAST \citep{devlin}, has
been carried out at these wavelengths. Models based on surveys at shorter wavelengths, such
as IRAS, are
likely to underestimate the number of sources in the submillimetre band because these surveys
miss the cold dust that radiates strongly at the longer wavelengths \citep{d2001}. 

We 
have used
two models to predict the properties of the sources that will
be found in the H-ATLAS. First, the model of Lagache and collaborators
\cite{guilaine1,guilaine2} is an empirical evolution
model in which an analytical form is assumed for the evolution, and the parameters of the
model are
adjusted to fit
the observational data in the far-infrared and submillimetre
wavebands, in particular the source counts at 24, 70, 160 and 850 $\mu$m and the spectrum of
the cosmic background radiation. 
Second, we have developed
a new empirical evolution model based on the SCUBA Local Universe and Galaxy Survey \cite{dunne2000,cat2005}.
Both models give adequate agreement
with the results of the BLAST survey \citep{dye2009}, although the comparison
between the results and the models is still at a preliminary stage because
of the large amount of source confusion in the BLAST survey \citep{eales09}. We now 
describe the new model in more detail. 

The model is based on the sample of 104 galaxies observed by Dunne et al. (2000) with the
SCUBA submillimetre camera as part of the SCUBA Local Universe and Galaxy Survey (SLUGS). These
galaxies form a statistically-complete sample above a flux limit 
at 60 $\mu$m of 
5.24 Jy \cite{dunne2000}. All of these galaxies were detected at 850 $\mu$m and many at
450 $\mu$m. Dunne and Eales (2001) present a simple two-component dust model that
fits the 60, 100, 450 and 850 $\mu$m flux measurements. 
For the galaxies without measurements at 450 $\mu$m, it is possible to determine
the parameters of the model by making the additional assumption that the
temperature of the cold dust component is 20 K, which is the average of the estimates
for the galaxies that do have complete flux measurements.
The SLUGS sample is still the only large
sample of galaxies for which there are empirical spectral energy distributions that
extend from the far-infrared to submillimetre waveband. 
More sophisticated attempts to use the SLUGS data to predict the local luminosity function
suggest that the simple method we present here leads to an underestimate
of the local luminosity function \citep{cat2005}. 
However, the local submillimetre luminosity function
is sufficiently uncertain that we are not too concerned about this---determining 
the luminosity function is one of the
goals of the H-ATLAS---and the simple modelling method we present below has
the advantage
that it is possible to make predictions for the optical, radio and other
properties of the galaxies we will detect with H-ATLAS.

We can use the SLUGS sample to predict 
the source counts in any submillimetre waveband in a straightforward way.
Let us, for example, predict the number of sources in the SPIRE 250 $\mu$m band. If we assume `number-density evolution',
in which the number of sources of a given luminosity changes with redshift, the number of sources
above a given 250 $\mu$m flux density is
\bigskip
$$
N(>S_{250 \mu m}) = \sum_{j=1}^{104} \int_0^{z(L_j,S_{250 \mu m})} {E(z) \over V_j} dV \eqno(1)
$$

\bigskip
\noindent in which $V_j$ is the comoving volume in which the j'th SLUGS source could have been detected
in the original survey from which it was selected and $E(z)$ is the ratio of the comoving number-density
of sources of a given luminosity at a redshift $z$ to the number at zero redshift. 
The upper limit of the integral is the maximum redshift at which the j'th source could be placed and
just be detected above the 250-$\mu$m flux limit.
$L_j$ is the
luminosity of the j'th SLUGS source, which can be estimated at the appropriate rest-frame wavelength at each redshift
(for the observed wavelength of 250 $\mu$m) from the two-component dust model.
In the specific case of the submillimetre waveband, we can use the two-component dust model \cite{d2001}
to estimate the luminosity at any wavelength. If we assume that rather than density-evolution we have
luminosity evolution, in which the number of galaxies stays the same but their 
luminosities evolve, the equation becomes:
\bigskip
$$
N(>S_{250 \mu m}) = \sum_{j=1}^{104} \int_0^{z(L_j,S_{250 \mu m})} {1 \over V_j} dV \eqno(2)
$$
\bigskip
\noindent In this case, all the information about cosmic evolution is included
in the upper limit because
\bigskip

$$
L_j(z) = E(z) L_j(0) \eqno(3)
$$

\bigskip
\noindent in which $E(z)$ is the ratio of the luminosity of a galaxy at redshift $z$ to
the luminosity of the galaxy at zero redshift.
These equations can be modified in a straightforward way to estimate the source counts
at any submillimetre wavelength and also the
cosmic background radiation.

In practice, we have used the simple luminosity-evolution model from Rowan-Robinson (2001),
in which the rest-frame monochromatic luminosity over the entire far-IR/submillimetre
waveband is assumed to evolve in the following way:
\bigskip
$$
L(t) = L(t_0) \left( {t \over t_0} \right)^P e^{Q(1-\frac{t}{t_0})} \eqno(4)
$$
\bigskip
\noindent in which $t$ is the time from the big bang and $t_0$ is the time at the
current epoch. $P$ and $Q$ are parameters of the model, and we found that 
$P=3$ and $Q=9$ produced acceptable fits to the spectral shape and intensity of the
cosmic background radiation \cite{fix98} and to the SCUBA 850 $\mu$m and
Spitzer 70$\mu$m source counts (Coppin et al. 2006, Frayer et al. 
2006a,b).

\begin{deluxetable}{lll}
\tabletypesize{\scriptsize}
\tablecaption{Predictions}
\tablewidth{0pt}
\tablehead{
\colhead{Wavelength/redshift range} &  \colhead{SLUGS} & \colhead{Lagache} \\
}
\startdata
110 $\mu$m, total & 76944 & 65973 \\
110 $\mu$m, $z<0.1$ & 15489 & 7267 \\
110 $\mu$m, $z<0.3$ & 39695 & 18502 \\
170 $\mu$m, total & 68740 & 56351 \\
170 $\mu$m, $z<0.1$ & 10088 & 7049 \\
170 $\mu$m, $z<0.3$ & 26761 & 17883 \\
250 $\mu$m, total & 222061 & 170783 \\
250 $\mu$m, $z<0.1$ & 12500 & 13321\\
250 $\mu$m, $z<0.3$ & 40073 & 34887\\
350 $\mu$m, total & 50422 &  32636\\
350 $\mu$m, $z<0.1$ & 2967 & 3192\\
350 $\mu$m, $z<0.3$ & 7389 & 8170\\
500 $\mu$m, total & 9734 & 6998\\
500 $\mu$m, $z<0.1$ & 856 & 444\\
500 $\mu$m, $z<0.3$ & 1714 & 1118\\
\enddata
\tablecomments{Reading from the left, the columns are:
the wavelength and the redshift range (`total' means all redshifts);
the prediction of the model based
on the SCUBA Local Universe and Galaxy Survey (see
text for details);
the prediction of the Lagache model \citep{guilaine1,guilaine2}}
\end{deluxetable}

Table 2 shows the total number of sources that this model and the models of Lagache
et al. (2003,2004) predict should be detected by H-ATLAS, including
the number at $\rm z < 0.1$ and at $\rm z < 0.3$. Figure 3 shows
the redshift distributions predicted by the two models. 
The numbers of sources predicted by the models at the
five wavelengths agree fairly well, although there are large differences
between the predicted redshift distributions. This difference is not surprising
because the observational data used to constrain the models consists almost entirely
of number counts, which inform us about the numbers of sources in slices of the
luminosity-redshift plane but nothing about the distribution of redshifts
within each slice.
Both sets of models do agree, however,
in predicting that the H-ATLAS will detect a large number of sources in the
relatively nearby Universe ($\rm z < 0.3$).

The advantage of the models based on SLUGS is that it is possible to predict
the properties of the H-ATLAS sources in any waveband in which the spectral energy
distributions (SED) of the SLUGS galaxies have been measured. For example,
the number of H-ATLAS galaxies that are predicted to have B-band optical magnitudes
in the range $\rm B_1 < B < B_2$ is given by
\bigskip
$$
N(B_1 < B < B_2) = \sum_{j=1}^{104} \int_{z_{min}(B_1,B_2,S_{250 \mu m},L_j)}^
{z_{max}(B_1,B_2,S_{250 \mu m},L_j)}
 {1 \over V_j} dV \eqno(5)
$$
\bigskip
\noindent in which the limits of the integral are the maximum and minimum redshifts
at which the j'th galaxy would fall both within the optical limits and above the
250-$\mu$m limit. In calculating these limits, we have assumed that the far-IR/submillimetre luminosity of the SLUGS galaxy
is evolving with cosmic time
in the way given by equation 4 but that its luminosity in the other
waveband is not evolving. This assumption is almost certainly not correct, but given our lack of knowledge
of the cosmic evolution in most wavebands it seems the safest (and simplest) one to make. It is also
a conservative assumption, in the sense that the model probably underestimates how bright the
H-ATLAS sources will be in the other wavebands. 
The SED of each galaxy is necessary to relate the flux at the observed wavelength---in this case
the B-band---to the luminosity of the galaxy at the wavelength at which the radiation was
emitted.
Similar equations can
be written for any waveband. 

Figures 4-8 show the predictions for the H-ATLAS galaxies in five bands. The panel
on the left-hand side shows the prediction for all the H-ATLAS galaxies and the
one on the right for those galaxies at $\rm z < 0.3$.

\begin{figure*}
\figurenum{4}
\plotone{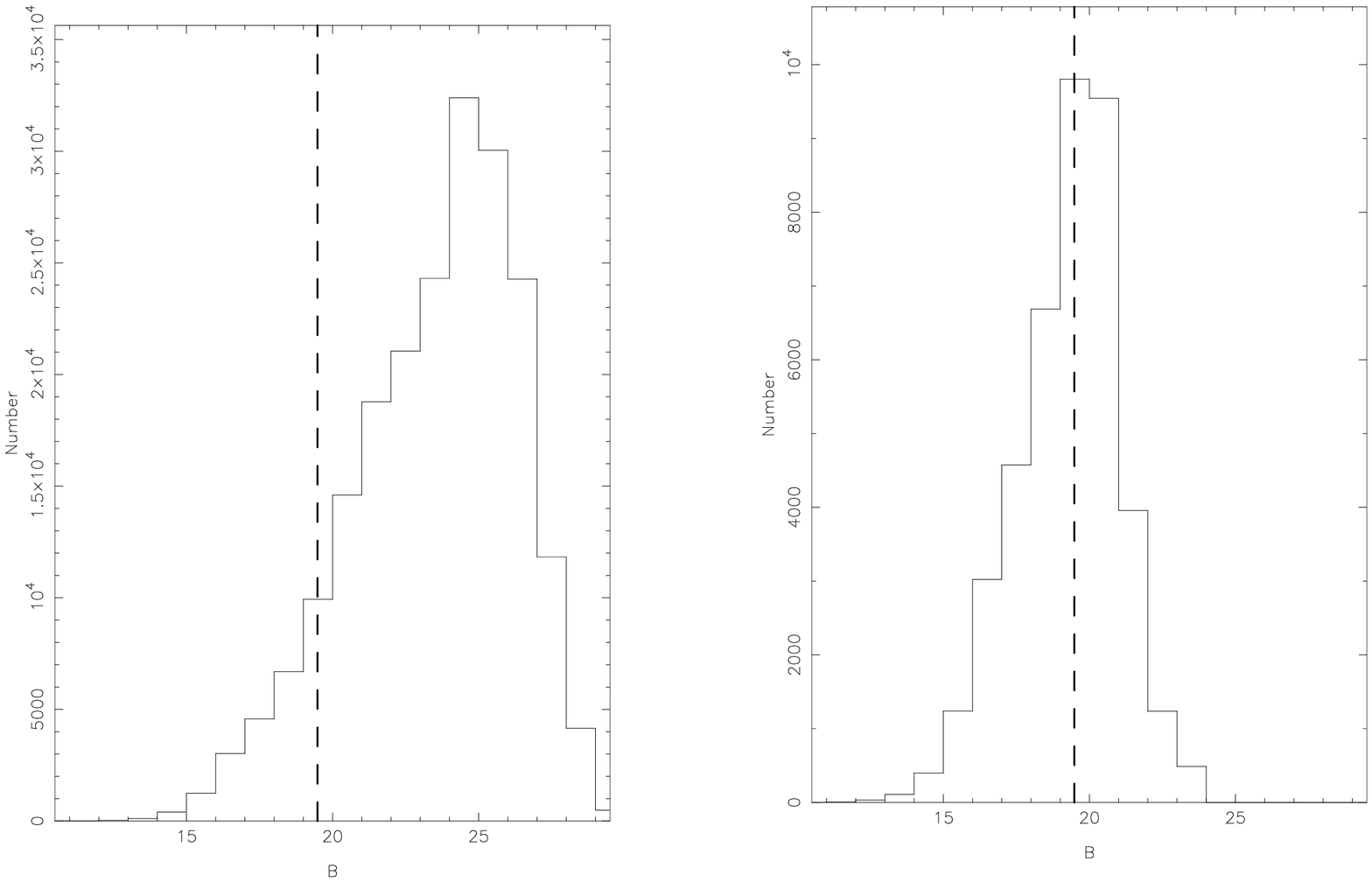}
\caption{The histogram of B-band optical magnitude predicted for the H-ATLAS galaxies using the
method described in the text. On the left is the prediction for all the galaxies and on the right
is the prediction for those at $\rm z < 0.3$. The vertical dashed line shows
the magnitude limit of the 2dF Galaxy Redshift Survey.}
\end{figure*}

\begin{figure*}
\figurenum{5}
\plotone{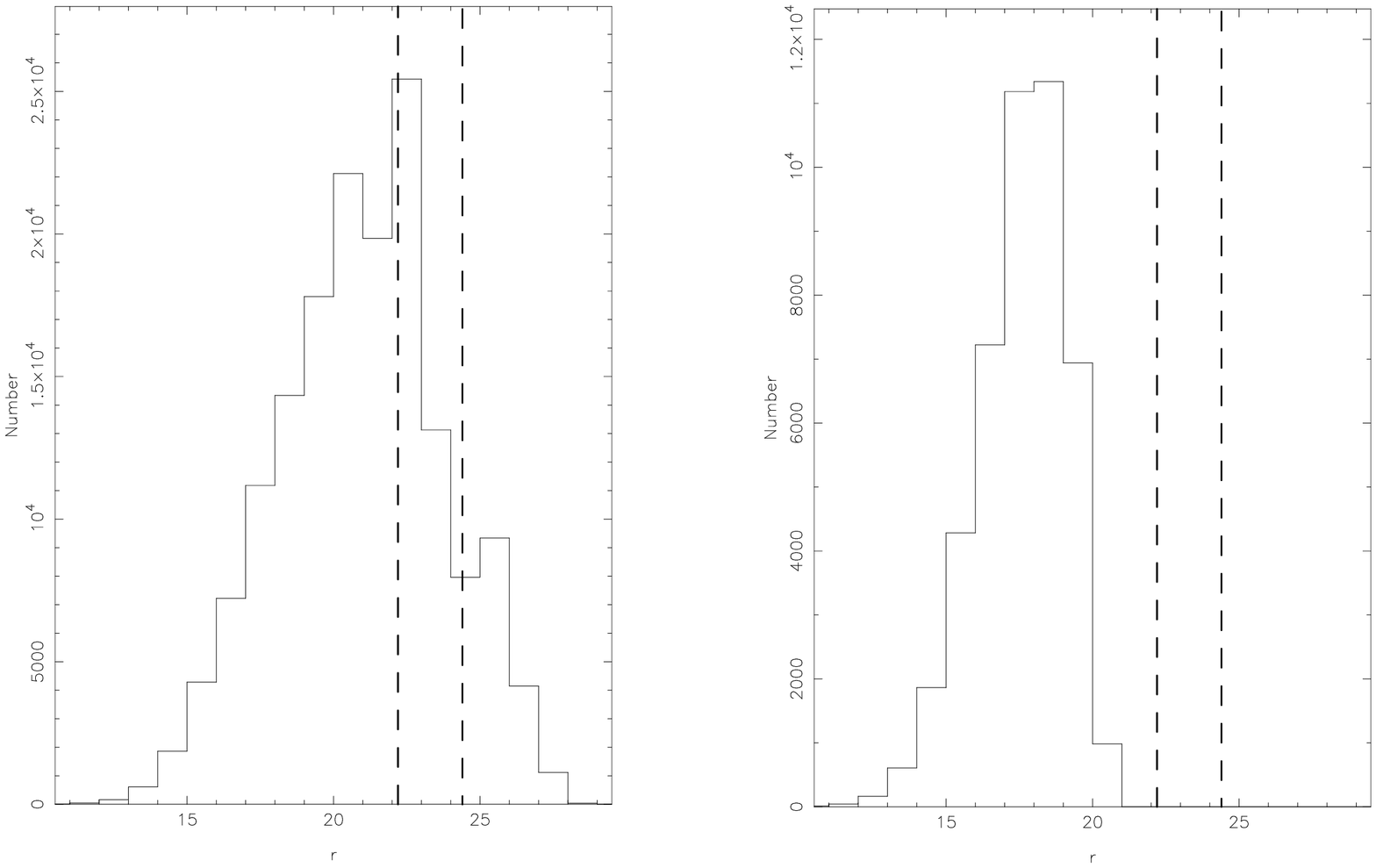}
\caption{The histogram of r-band optical magnitude predicted for the H-ATLAS galaxies using the
method described in the text. On the left is the prediction for all the galaxies and on the right
is the prediction for those at $\rm z < 0.3$. The
vertical dashed line on the left shows the limiting magntiude
of the imaging part of the Sloan Digital Sky Survey; the one on the right shows
the approximate limit that will be reached by the ESO public survey KIDS.}
\end{figure*}

\begin{figure*}
\figurenum{6}
\plotone{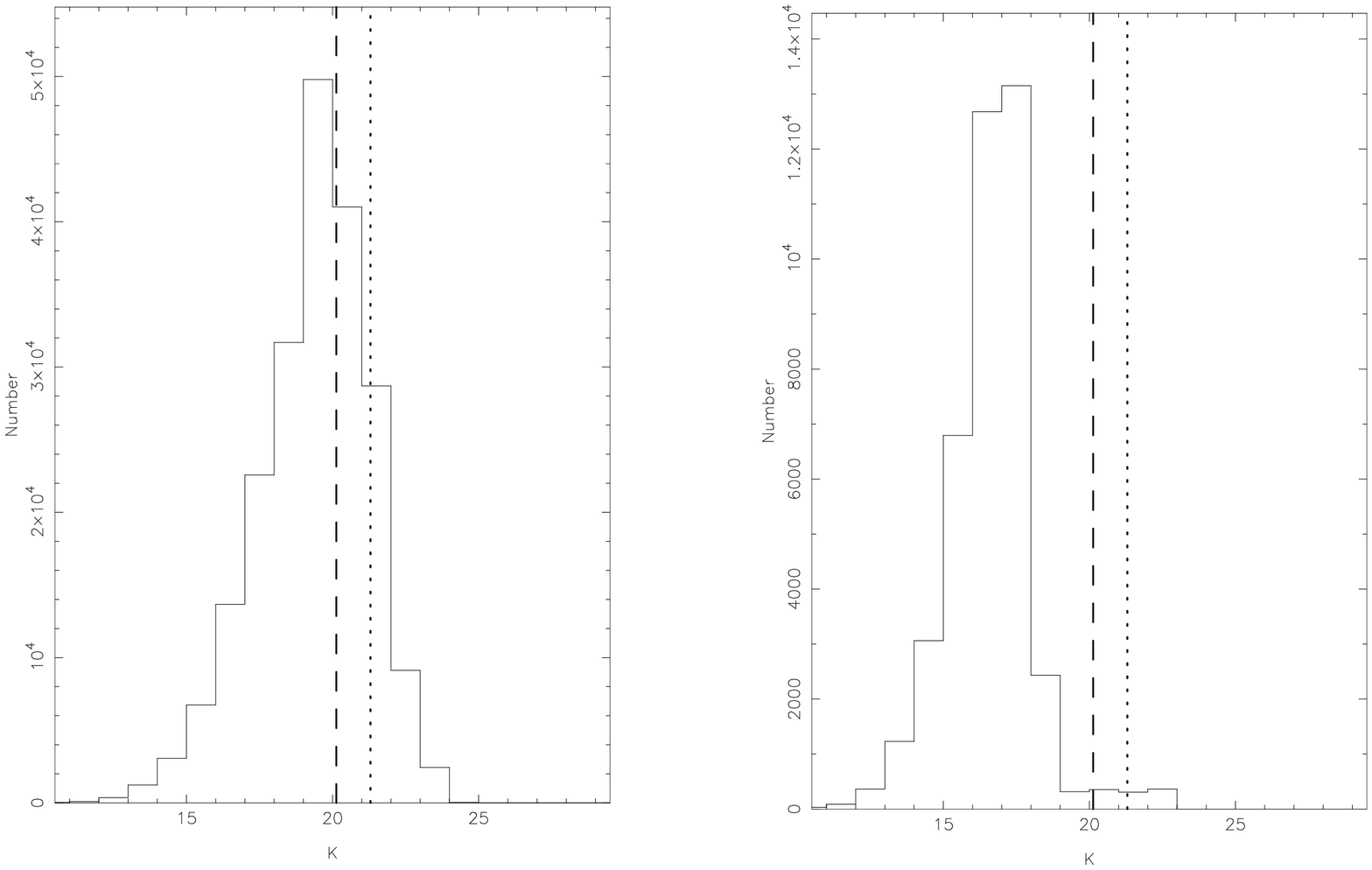}
\caption{The histogram of K-band optical magnitude predicted for the H-ATLAS galaxies using the
method described in the text. On the left is the prediction for all the galaxies and on the right
is the prediction for those at $\rm z < 0.3$. The left-hand vertical dashed line shows the limiting magntiude
of the Large Area Survey that is part of the UKIRT Infrared Deep Sky Survey \cite{stevew}. The right-hand vertical
line shows the limiting magnitude of the ESO public survey, VIKING.}
\end{figure*}

\begin{figure*}
\figurenum{7}
\plotone{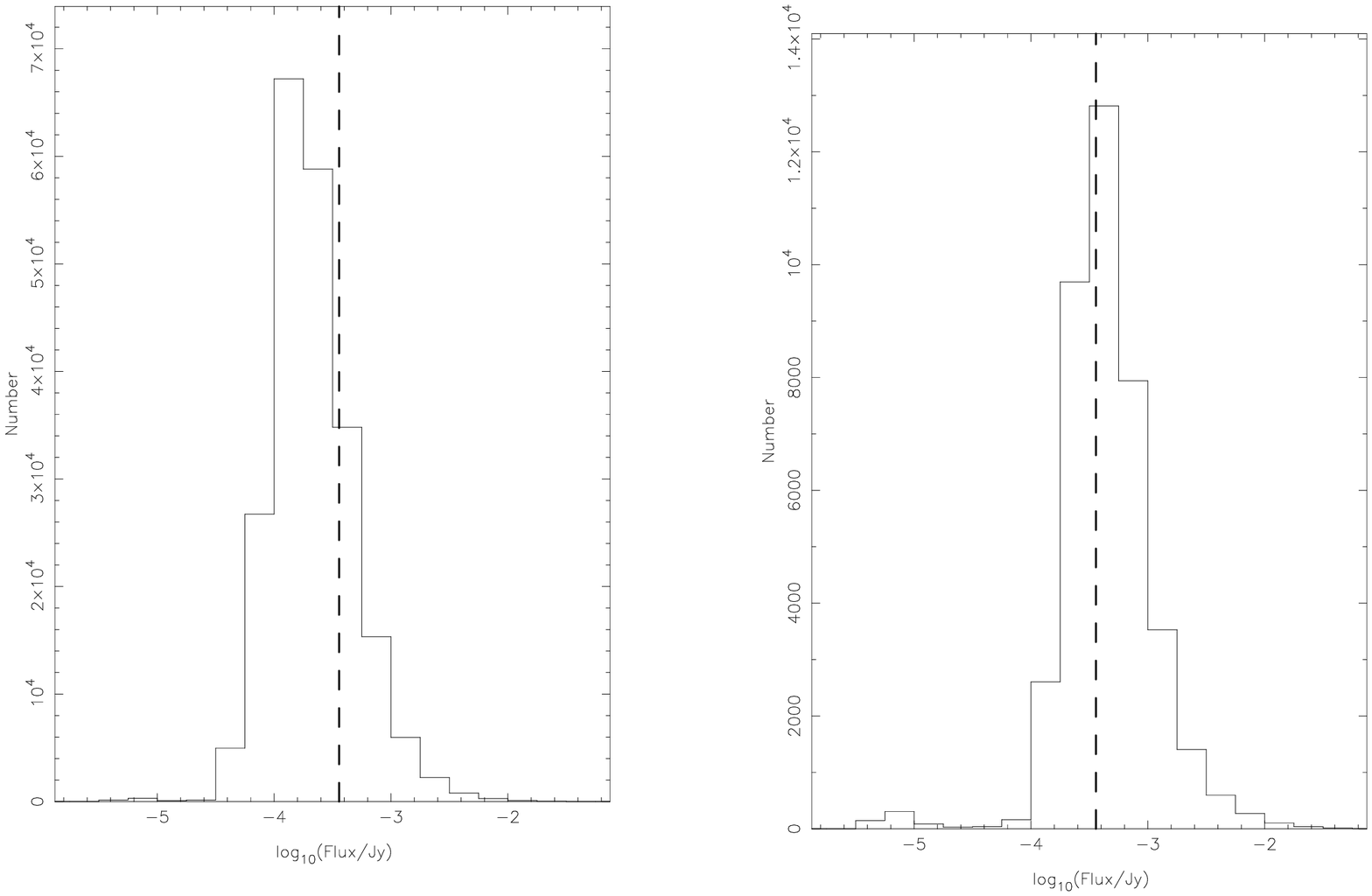}
\caption{Histogram of the 1.4-GHz radio fluxes predicted for the H-ATLAS galaxies using the
method described in the text.
On the left is the prediction for all the galaxies and on the right
is the prediction for those at $\rm z < 0.3$. The vertical dashed line shows the approximate
5$\sigma$ limit of the radio survey we are
carrying out with the GMRT, translated to 1.4 GHz using the assumption of a power-law radio
spectrum ($\rm S \propto \nu^{-\alpha}$) with $\rm \alpha=0.7$ (\S 4.3).}
\end{figure*}

\begin{figure*} 
\figurenum{8} 
\plotone{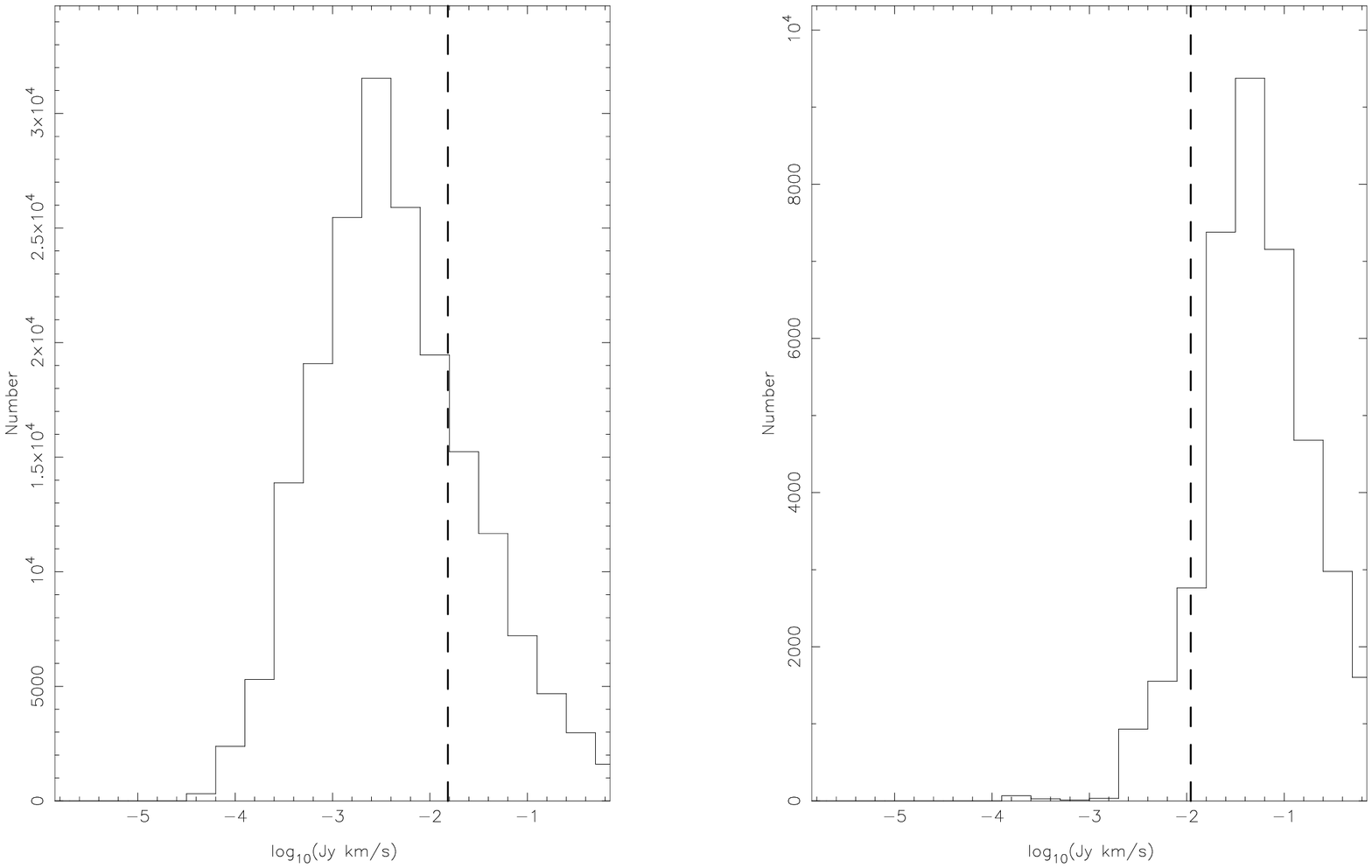}
\caption{Histogram of the 21-cm line fluxes predicted for the H-ATLAS galaxies using the
method described in the text.
On the left is the prediction for all the galaxies and on the right
is the prediction for those at $\rm z < 0.3$. The vertical dashed line shows the approximate
5$\sigma$ limit of the HI survey described in \S 4.3.}
\end{figure*}

Figures 4 and 5  show the predictions for the optical B and r bands. 
Several major redshift surveys have been based on catalogues defined in these
bands and in \S 4 we use these predictions to estimate the fraction of the H-ATLAS galaxies
that will already have spectroscopic redshifts. 
In making the predictions for the
B band, we have used the B-band photometry that exists in the NASA Extragalactic Database
(NED) for most of the SLUGS galaxies, and we made the
assumption that all the SLUGS galaxies have an SED typical
of an Sbc galaxy \cite{cole2}. 
Note that although the assumption of a single optical SED for all
H-ATLAS galaxies is a crude one, it should introduce little error at low redshift, where the observed and rest frames
are very close in wavelength. We have made the r-band predictions using the V-band photometry, which exists in NED
for 75\% of the SLUGS galaxies, with a small correction to the r-band using the colour transformations in
Jester et al. (2005). 

Figure 6 shows the predictions for the K-band. We have used the K-band photometry in NED 
that exists for
almost all the SLUGS galaxies (mostly from 2MASS) and again assumed a standard Sbc SED.

Figure 7 shows the predicted 1.4-GHz continuum radio fluxes of the H-ATLAS
galaxies. We have made the prediction for
the continuum radio fluxes using the 1.4-GHz flux measurements that exist for almost all the SLUGS
galaxies plus the assumption that all the SLUGS galaxies have a power-law radio continuum ($S \propto
\nu^{-\alpha}$) with $\alpha=0.7$. In this case, it seems quite likely that the assumption of
no evolution in the radio waveband is incorrect because of the strong correlation between the
far-infrared/submllimetre and radio emission from star-forming galaxies both at low and
and high redshift \cite{helou,ibar}. We have therefore assumed that radio luminosity also evolves in the way given
by equation 4.

Figure 8 shows the prediction for the HI line fluxes of the H-ATLAS galaxies. We used
the HI line fluxes for the SLUGS galaxies given in NED. We have assumed that the amount of
atomic hydrogen in a galaxy does not evolve, which may mean that the predicted HI line
fluxes of the H-ATLAS galaxies are underestimates.

The figures show that, in contrast to the sources detected in the SCUBA surveys at 850 $\mu$m, it should be
fairly easy to follow-up the H-ATLAS sources with observations in other wavebands.

\section{Multi-Wavelength Data}

We selected our fields partly because of the low `cirrus emission' from dust in our
own galaxy (Fig. 2) but mainly because the complementary multi-wavelength data  
is better than for any field of similar size. We describe here the mult-wavelength datasets,
both ones that exist now and ones that are likely to soon exist. 

\subsection{Spectroscopy}

There have been three major recent redshift surveys: the Sloan Digital Sky Survey (SDSS - York
et al. 2000);
the 2dF Galaxy Redshift Survey (2dFGRS - Colless et al. 2001) and
the Galaxy And Mass Assembly Survey (GAMA - Baldry et al. 2009,
Driver et al. 2009). The GAMA survey
is still underway. All of the H-ATLAS fields are covered by one or more
of these surveys. We have used the models from the last section to predict
the number of H-ATLAS sources that will already have redshifts from one of these
surveys. In making these predictions, we have made the assumptions that (a) the
2dFGRS measured redshifts for 93\% of the galaxies with $B < 19.6$ \citep{colless} 
\footnote{We have made the trasnformation
from the $b_j$ magnitude system using the relationship $b_j = B - 0.28(B-V)$ given
in Maddox et al. (1990).}, (b) that the SDSS measured redshifts for 94\% of galaxies
with $r < 17.77$ \citep{strauss} and (c) that GAMA will measure redshifts for all
galaxies with $r < 19.4$ \citep{driver}. Table 3 lists the predicted numbers of sources which will
already have spectroscopic redshifts in the different fields.
The table shows that $\sim2\times10^4$ H-ATLAS galaxies are likely to already have spectroscopic redshifts,
and that the percentage of H-ATLAS galaxies at $z < 0.1$ with spectroscopic redshifts is likely to 
be very
high. All three surveys, of course, contain far more information about the 
galaxies than only the  redshifts, such as measurements of line ratios, line
equivalent widths, kinematics etc.

\begin{deluxetable*}{llll}
\tabletypesize{\scriptsize}
\tablecaption{Redshift Surveys}
\tablewidth{0pt}
\tablehead{
\colhead{H-ATLAS field} & \colhead{Datasets} & \colhead{No. at $z < 0.1$} &  \colhead{No. at $z < 0.3$} \\
}
\startdata
 NGP & SDSS & 3366 (2895) & 10865 (4563)   \\
GAMA & GAMA,SDSS,2dFGRS & 3366 (3366) & 10865 (9456) \\
SGP & 2dFGRS & 5610 (4993) & 18106 (8148)   \\
All fields & ... & 12342 (11254) & 39833 (22167 ) \\
\enddata
\tablecomments{Reading from the left, the columns are:
the H-ATLAS fields; the redshifts surveys covering this field; the predicted number
of sources at $z < 0.1$ in this field with the predicted number with spectroscopic redshifts
in brackets;
the predicted number
of sources at $z < 0.3$ in this field with the predicted number with spectroscopic redshifts
in brackets}
\end{deluxetable*}

\subsection{Imaging from the ultraviolet to the near infrared}

At ultraviolet wavelengths, approximately half the total survey area has been observed with GALEX. This has
mostly been part of the GALEX Medium Imaging Survey (MIS), which has an exposure time of 1500s and
a limiting AB magnitude of $\simeq$23.
We and the GAMA team have been awarded
time for a proposal to complete the GALEX coverage of the GAMA fields to
the MIS depth (P.I. Tuffs). 

In the optical waveband, the GAMA fields and the NGP field have been surveyed in five passbands
by the Sloan Digital Sky Survey. The limiting magnitude of the r-band imaging is shown
in Figure 5. As yet, the SGP field has only been 
surveyed with the UK Schmidt Telescope using photographic plates. 
In the next three years, the H-ATLAS fields should be covered by three new optical
surveys.
The GAMA and 
SGP fields will be observed in four passbands
as part of the Kilo Degree Survey (KIDS), an ESO public 
survey that will be carried out the with VLT Survey Telescope (VST). The sensitivity limits
of KIDS will be about 2 magnitudes fainter than the SDSS limits.
The NGP and GAMA fields will be observed in five bands by Pan-STARRS1 and the SGP and GAMA fields
will be observed in six bands by SkyMapper \cite{sky}. The approximate 5$\sigma$ limits
for all the optical surveys are given in Table 4. A little further off in time, the SGP field
will be observed as part of the Dark Energy Survey.

In the near infrared, the GAMA and NGP fields are being surveyed in four passbands as part of the
Large Area Survey (LAS), a legacy survey being carried out as part of
the UKIRT Infrared Deep Sky Survey \cite{stevew}. The GAMA and SGP fields will soon be observed in five passbands
as 
part of the VISTA Kilo-Degree Infrared Galaxy Survey in the Infrared (VIKING), an ESO
public survey that is currently
being carried out with the Visible and Infared Telescope for Astronomy (VISTA).
The approximate limits of both surveys are given in Table 4 and shown in Figure 6.

\begin{deluxetable*}{llllllllllllll}
\tabletypesize{\scriptsize}
\tablecaption{Optical and Near Infrared Data}
\tablewidth{0pt}
\tablehead{
\colhead{H-ATLAS field} &  \colhead{Datasets} & \colhead{u}  & \colhead{v} & \colhead{g} &
\colhead{r} & \colhead{i} & \colhead{z} & \colhead{Z} &
\colhead{y} & \colhead{Y} & \colhead{J} & \colhead{H} &
\colhead{K}\\
}
\startdata
NGP  & SDSS & 22.0 & ... & 22.2 & 22.2 & 21.3 & 20.5 & ... & ... & ... & ... & ... & ... \\
NGP & Pan-STARRS1\tablenotemark{a} & ... & ... & 24.1 & 23.5 & 23.4 & 22.4 & ... & 21.2 & ... & ... & ... & ... \\
NGP & LAS & ... & ... & ... & ... & ... & ... & ... & ... & 20.87 & 20.55 & 20.28 & 20.13 \\
GAMA & SDSS & 22.0 & ... & 22.2 & 22.2 & 21.3 & 20.5 & ... & ... & ... & ... & ... & ... \\
GAMA & KIDS & 24.0 & ... & 24.6 & 24.4 & 23.4 & ... & ... & ... & ... & ... & ... & ... \\
GAMA & Pan-STARRS1\tablenotemark{a} & ... & ... & 24.1 & 23.5 & 23.4 & 22.4 & ... & 21.2 & ... & ... & ... & ... \\
GAMA & SkyMapper\tablenotemark{b} & 22.9 & 22.7 & 22.9 & 22.6 & 22.0 & 21.5 & ... & ... & ... & ... & ... & ... \\
GAMA & LAS & ... & ... & ... & ... & ... & ... & ... & ... & 20.87 & 20.55 & 20.28 & 20.13 \\
GAMA & VIKING & ... & ... & ... & ... & ... & ... & 23.1 & ... & 22.4 & 22.2 & 21.6 & 21.3 \\
SGP & KIDS & 24.0 & ... & 24.6 & 24.4 & 23.4 & ... & ... & ... & ... & ... & ... & ... \\
SGP & SkyMapper\tablenotemark{b} & 22.9 & 22.7 & 22.9 & 22.6 & 22.0 & 21.5 & ... & ... & ... & ... & ... & ... \\
SGP & VIKING & ... & ... & ... & ... & ... & ... & 23.1 & ... & 22.4 & 22.2 & 21.6 & 21.3 \\
\enddata
\tablenotetext{a}{The limits are for the planned three-year survey with Pan-STARRS1.}
\tablenotetext{b}{The limits are for the planned six-epoch survey by SkyMapper.}
\tablecomments{Reading from the left, the columns are:
the H-ATLAS field; the optical or near-IR survey; the quoted sensitivity limits
in
AB magnitudes (These are 5$\sigma$ limits for point sources, except in the case of the SDSS in which the limits represent the 95\% detection repeatability for point sources).}
\end{deluxetable*}

\subsection{Radio Surveys - continuum and HI}

The H-ATLAS fields have been surveyed at 1.4 GHz as part of the NRAO VLA Sky Survey. This survey is not
sensitive enough, however, to detect a significant percentage of the H-ATLAS sources. We
are therefore currently completing a radio survey (P.I. Jarvis) of the GAMA 
fields at 325 MHz with the Giant Metre-wave
Radio Telescope (GMRT), which will reach an approximate 5$\sigma$ sensitivity of
1 mJy. The limit of this survey is shown in Figure 7. 
The NGP field is also
high on the list of targets for an early survey with the Low Frequency Array 
for Astronomy (LOFAR).

There is currently no HI survey of any of the H-ATLAS fields that could yield HI measurements
of a significant fraction of the H-ATLAS sources. However, the SGP and GAMA fields are natural
targets for the new southern radiotelescopes that are being built as prototypes for the Square
Kilometre Array (and for the SKA itself, of course). 
We have designed a potential HI survey of the SGP field using the parameters given
for the Australian Square Kilometre Pathfinder Array (ASKAP) by Johnston et al. (2007).
We estimate that in one month of observing time it should be possible to carry out a survey
of the SGP sensitive to galaxies out to $\rm z \simeq 1$ with an approximate sensitivity to HI flux
of $\rm \simeq 10^{-2}\ Jy\ km\ s^{-1}$ (Fig. 8).

\subsection{Planck and other telescopes}

The Planck Surveyor is surveying the whole sky in nine passbands, two of which (350 and
550 $\mu$m) are the nearly the same as those that will be used in H-ATLAS. 
The size (full-width-half-maximum) of the Planck beam is $\simeq10$\ times greater than that of H-ATLAS at
the same frequency, although 
the sensitivity in surface brightness is fairly similar.
Therefore, in the common area
of the sky (one eightieth of the whole sky), the two surveys will be complementary, with the
H-ATLAS providing high-resolution observations of the sources that will be detected by 
Planck.

The SGP field may also be observed by the South Pole Telescope, a telescope designed to
look for high-redshift clusters using the Sunyaev-Zeldovich effect.

\section{The H-ATLAS Science Programme}

In this section we describe the six major science programmes planned by the 
H-ATLAS team. We note that these programmes
represent only a limited subset of the scientific projects that will be ultimately possible with the H-ATLAS for
two reasons. First, many other projects will become possible as the surveys of the H-ATLAS fields at other
wavelengths (\S 4) are gradually completed. Second, since the wavelength range from 200 to 500 $\mu$m is virtually
unexplored and since the H-ATLAS will cover one eigthieth of the sky in this waveband, it is possible that there
will be some unanticipated discoveries.

\subsection{The Local Universe}

Our models (\S 4)  predict that the H-ATLAS will detect $\simeq$40,000 galaxies in the relatively nearby Universe
($\rm z < 0.3$).
Almost all the galaxies detected at $z < 0.1$ and approximately
half the galaxies detected at $\rm z < 0.3$ should already have spectroscopic redshifts. Most of these galaxies
will be unresolved by the H-ATLAS beams.
There are 
$\simeq$120 clusters of Abell richness class 1 or greater, including the Coma Cluster, in our fields.

Apart from providing the spectroscopic redshifts necessary to calculate
the intrinsic properties of many of the sources (luminosities, dust masses etc.),
the complementary data will be helpful in other ways. 
Three are particularly important. First, the redshift surveys will allow us to determine the position within the
cosmic web of each
Herschel galaxy and to measure the density of the galaxy's environment (exactly how to do this best is
debateable---e.g. Balogh et al. 2004). Second, the existence of catalogues at other wavelengths means that the
H-ATLAS will contain more information than the individual source detections. By coadding (`stacking') the Herschel
emission at the positions of objects in different classes, we can study the dust and dust-obscured star formation
in objects that would be too faint to detect individually \citep{dole,dye2007}. An important 
example is elliptical galaxies (\S 1).
Third, we will be able to answer the question of what fraction of the optical light from a galaxy is
obscured by dust. Optical astronomers have struggled with this issue for 50 years \citep{holmberg,davies}
with limited success, largely because optical galaxy catalogues have large selection effects caused by dust. The
implications for extragalactic astronomy are potentially large, with
a recent
study concluding
that the optical luminosity function is significantly altered by dust extinction
and that even
bulges suffer as much as 2 mag of extinction at certain
inclinations \citep{driver2}. The ultraviolet, optical and near-infrared data that exist for some
of the fields (\S 4) will allow us to address this issue by the simple energy-balance technique
of comparing the total dust emission to the total unobscured
starlight. The only assumption in this technique is that the absorption by the dust is isotropic. This is almost certainly
untrue but by averaging over many objects we can largely eliminate the effects of anisotropy
(Driver et al. 2008).

We give the following four projects as examples:

\begin{itemize} 

\item We will make the first accurate estimate of the local submillimetre luminosity
and dust-mass functions down to dust masses of $\rm \sim 10^{4.5}\ M_{\odot}$
(Fig. 9). As we will have Herschel measurements for $\simeq$10,000 galaxies at $\rm z < 0.1$,
we will be able to compare the luminosity and dust-mass functions for different classes
of galaxy (e.g. different Hubble types, high and low environmental density etc.). We will also be able
to extend the study of the luminosity and dust-mass functions to higher dimensions, for example by
examining how the space density of galaxies depends on both dust mass and stellar mass.

\item Using the results from the redshift surveys,
we will investigate how
the dust-obscured star formation depends on the local and large-scale
environment. This complements previous optical studies \cite{balogh04,kauffmann04},
because our observations will be much more sensitive to starbursts,
which may well have a different environmental dependence from
quiescent star formation.

\item We will investigate how the dust content of the Universe and
dust-obscured star formation has changed during the last three billion
years. This will finally follow up an important discovery from IRAS
that there is strong evolution in the luminosity function at a
suprisingly low redshift \citep{saunders}, a phenomenon which also may have been
seen in the SDSS \citep{loveday}. 

\item One of the strongest correlations in astronomy is that between the far-infrared and nonthermal radio emission of 
galaxies \cite{helou,ibar}. A widely-accepted explanation is that both the far-infrared emission and the radio
emission are ultimately caused by young stars, with the far-infrared emission being from the
dust heated by young stars and the radio emission being nonthermal emission from the relativistic
electrons generated in the supernova remnants produced when the stars reach the end of their lives. Nevertheless,
many of the properties of this relationship are hard to explain, in particular the approximate unity slope
when the two quantities are plotted on logarithmic axes \cite{cat2}. By investigating this relationship as a function of environment,
redshift, star-formation activity and other properties, we will try to understand better its fundamental cause. 

\end{itemize} 

\begin{figure*}
\figurenum{9}
\epsscale{0.8}
\plotone{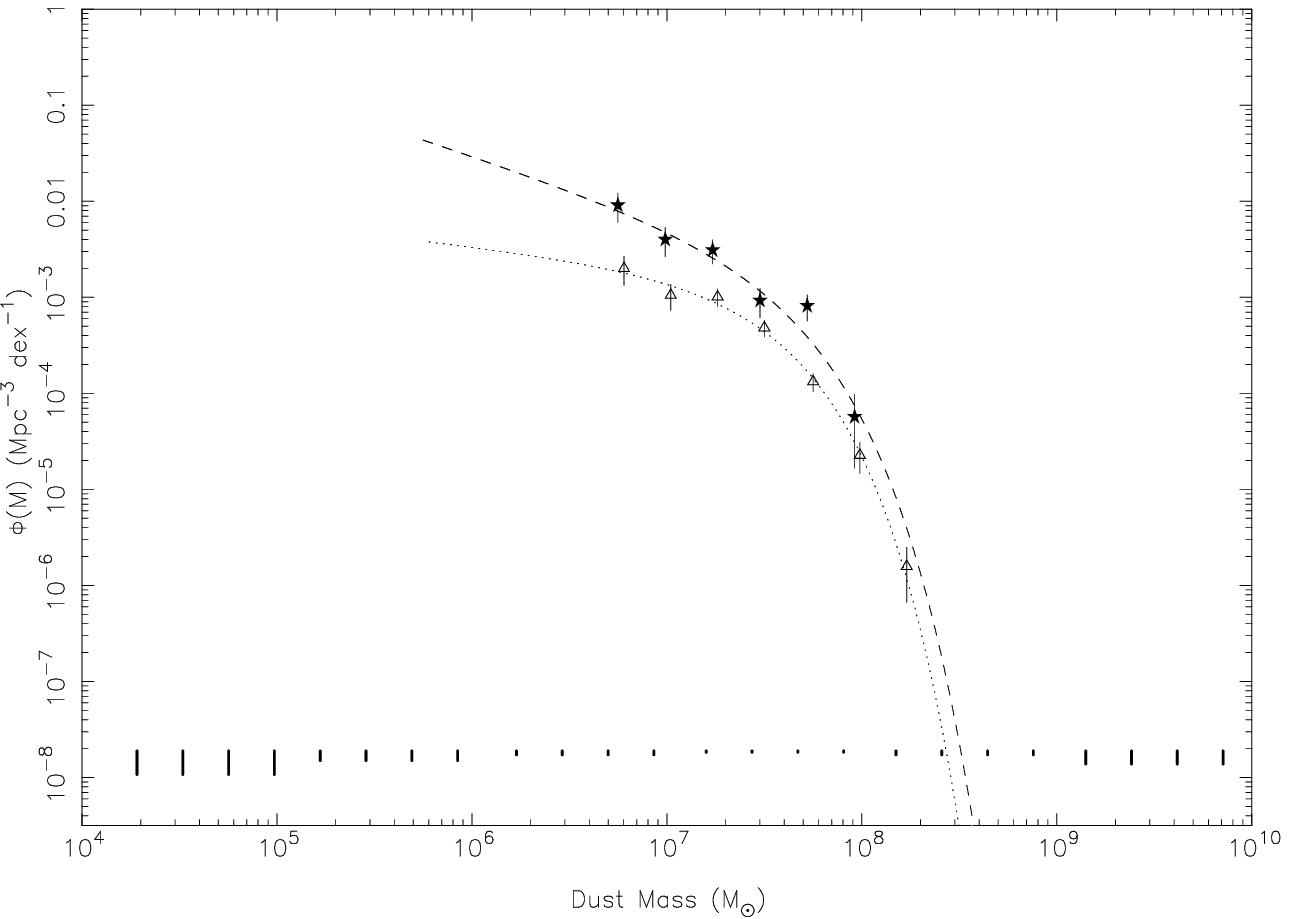}
\caption{The triangles and stars show estimates of the dust-mass function (the
space-density of galaxies as a function of dust mass) from Vlahakis, Dunne and Eales
(2005). The vertical lines along the bottom show our estimates, on the assumption
of Poisson statistics, of the $\pm1\sigma$ errors
on the estimate of the dust-mass function from the H-ATLAS.}
\end{figure*}

\subsection{Planck Point Sources and Diffuse Emission}

The Planck
High-Frequency Instrument (HFI) will survey the whole sky in six bands
(3, 2.1, 1.4, 0.85, 0.55, 0.35 mm), the first survey of the whole sky
at these wavelengths. Apart from its measurements of the primordial cosmic
background radiation (CMB), the HFI will detect extended foreground emission, such
as dust in the Galaxy, and it is also 
likely to detect many point sources \citep{dezotti}, including
nonthermal radio sources, thermal dust emission from nearby galaxies and 
Sunyaev-Zeldovich (SZ) sources associated with distant clusters,
the consequence of the scattering of the CMB radiation by the electrons in the intracluster medium.

The SZ sources detected by Planck are potentially of great 
importance for cosmologists, because the number-density of clusters as
a function of redshift depends critically on the cosmological model \citep{bart2001,bart2002}. It might also be
possible,
because the spectral shape of the SZ effect depends on the
peculiar motion of the cluster,
to measure bulk flows
in the universe \citep{kash}, which is another critical test of the cosmological
paradigm. A major problem, however, is
the contamination of the SZ effect by thermal emission from dusty galaxies
within the large (5-10 arcmin) Planck beam, because there is evidence that even
at moderate redshift
the combined emission from dust in cluster galaxies is comparable to the SZ effect in
the 0.85-mm band, and much greater in the two shorter wavelength bands \citep{zemcov}. There is
zero SZ effect at 1.4 mm,
which leaves only the two long-wavelength bands with little contribution from dusty galaxies,
but as these also have the worst angular resolution there is the additional problem of confusion with
nonthermal radio sources.

The H-ATLAS will survey one eightieth of the sky in the two Planck bands with the highest frequencies,
but with much better resolution. The sensitivity to extended regions of low surface brightness
of the two surveys is likely to be similar. The combination of the two surveys will make possible several joint projects:

\begin{itemize} 

\item The effects of confusion on the Planck point source catalogue are likely 
to be large, particularly in the highest frequency channels. These include Eddington bias \cite{edd}, which can lead to
both overestimates of the fluxes of the sources and spurious detections, and a large and uncertain effect on the positions
of the sources. The H-ATLAS will provide a sample of sources with accurately known positions and fluxes, which
can then be checked against the
Planck point-source catalogue for the same area of sky
for the  two highest frequency channels. This comparison
will provide an estimate of the correction factor for the Planck fluxes, an estimate of the fraction of the Planck sources that are spurious,
and measurements of the positional errors of the Planck sources. 

\item Some of the Planck science goals, such
as the detection of the anisotropies in the cosmic infrared background \cite{guilaine1,guilaine2},
require the successful removal of
galactic cirrus emission. The galactic cirrus should be easier to distinguish in the H-ATLAS images because of the
better angular resolution, and so we will be able to test the success
of the Planck component separation techniques.

\item The combination of the high-resolution images
provided by H-ATLAS and the spectral energy distributions of the Planck sources should make it possible
to determine unambiguously the nature of each source in the overlap region between the two surveys.
The detailed information provided by the H-ATLAS for the Planck sources in this region will make it
possible to develop better statistical methods based on their spectral energy distributions for
determining the nature of the Planck sources in the region not covered by H-ATLAS.

\item In the overlap region between the two surveys, the H-ATLAS
will make it possible to determine the contribution of dusty galaxies to each SZ source and then subtract
these galaxies' emission from the SZ signal. The detailed information in the overlap region will
also provide statistical information about the average level of contamination for the Planck SZ sources
not covered by H-ATLAS.

\end{itemize}

\subsection{The H-ATLAS Lens Sample}

In principle, gravitational lensing
is a powerful way
of investigating the evolution in the mass profiles of
galaxies, a fundamental test of models of structure formation.
In practice, it
has proved very hard to assemble the necessary large sample of lenses. The most ambitious programme to date
searched for lenses among flat-spectrum radio sources but,
after high-resolution
radio observations, found only 22 lenses out of 16000 radio sources---a
success rate of 0.14\% \citep{browne}.

\begin{figure*}
\figurenum{10}
\plotone{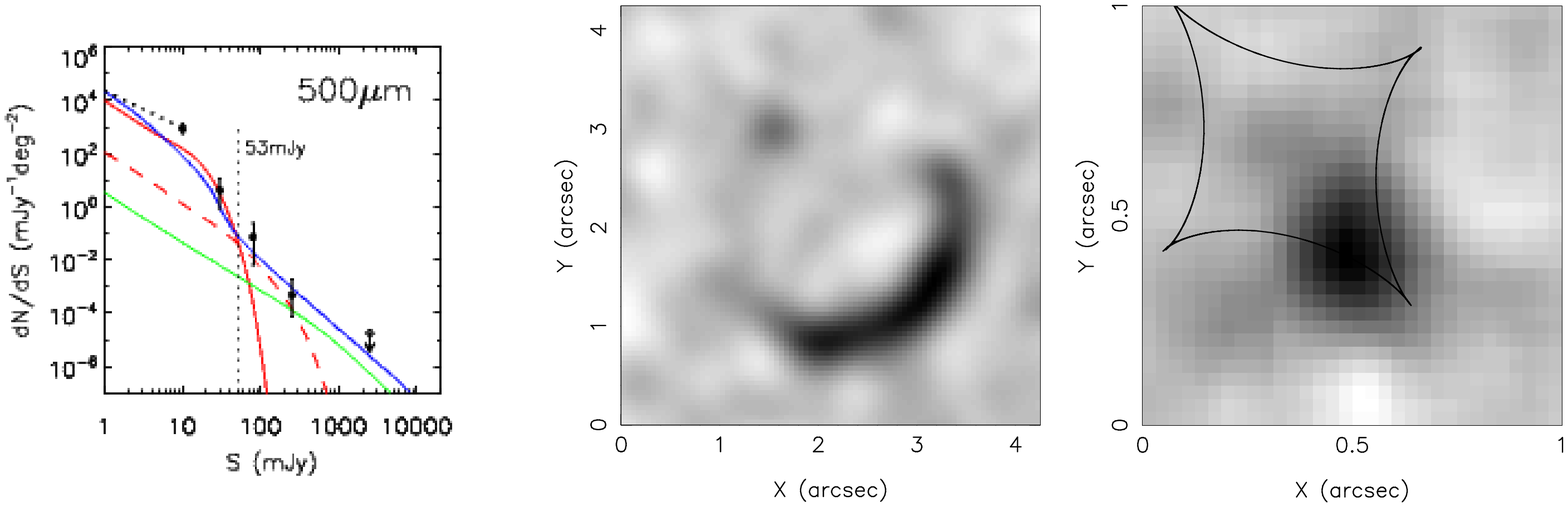}
\caption{The
left-hand panel shows the predicted
number of sources as a function
of 500 $\mu$m flux density plotted over the results from BLAST \citep{mattia}.
The key to the curves is as follows: red solid---unlensed high-$z$ protospheroids; red dashed---lensed
high-$z$ protospheroids; green---blazars; blue---nearby late-type galaxies.
The vertical dashed line shows the approximate limit of H-ATLAS at 500 $\mu$m.
The middle panel shows a simulated image of a strongly lensed
100mJy 500$\mu$m H-ATLAS source as observed using the Submillimetre Array at 870$\mu$m
in extended mode over a 6h track.  The panel on the right shows the
reconstructed source obtained by applying the reconstruction method of
Dye et al. (2007a) to the simulated ring image. The image caustic
of the best fit lens model is over-plotted.}

\end{figure*}

\hspace{0.3in}
Submillimetre surveys are possibly the ideal way to find
lenses. At $z \ge 1$, the monochromatic flux density at the long-wavelength end of the
submillimete waveband depends on luminosity but is approximately independent of redshift,
a consequence of the characteristic spectral energy distribution of dust in galaxies.
A result of this and the strong cosmic evolution in the submilimetre waveband is that sources in submillimetre surveys tend to lie at
high redshifts (\S 1) and are thus likely to have a large optical depth to lensing.
Because of the approximate independence of flux and redshift, a bright source in a submillimetre survey must
necessarily have a high luminosity unless it is is at $z \le 1$.
Above a critical flux density, the luminosities of any sources beyond this redshift must lie above the
luminosity at which the luminosity function declines steeply---and so any sources above this flux density
are highly likely to be lensed systems.

The model in Figure 10, for example, predicts that at
$\rm S_{500 \mu m} > 100\ mJy$ the sources should be a mixture of lensed high-redshift galaxies, nearby galaxies
and flat-spectrum radio sources \citep{mattia}. Since the latter two catagories would be easy to remove (by using the
submillimetre flux ratio or the presence of a bright galaxy), the lens yield
of such a sample would be close to 100\%.
These models predict that the H-ATLAS will contain $\sim 1500$, 800 and 350 strongly-lensed
galaxies at 250, 350 and 500$\mu$m, respectively.

This is the one of the H-ATLAS science programmes for which follow-up observations will be necessary, both to
determine whether a bright 500-$\mu$m source is indeed a lensed system  and to scientifically exploit the 
H-ATLAS lens
sample. Fortunately, a source with
$\rm S_{500 \mu m} \sim 100\ mJy$
would be be easy to map; a 30-minute observation with
the Submillimetre Array, a one-hour observation with the VLA or a $\simeq$1-min observation with
ALMA
would be enough to confirm a source is a lens and to map the image structure.
The prediction that large-area submillimetre surveys are the ideal way to construct large
samples of lenses may not be correct, but if it is, there are many possible uses
for a large sample of lenses. These include: 

\begin{itemize} 

\item an investigation of the evolution of the profiles of the lenses (probably mostly
elliptical galaxies). In principle, it should be possible to use standard techniques \citep{simon3}
to reconstruct separately the structures of the dark-matter halo and baryonic component of each
lens.

\item a study of the structures of high-redshift dust sources by reconstructing the original (unlensed) structure
of the source (Fig. 10). Since our systems are lensed, we will be able to study galaxies well below the
Herschel confusion limit \cite{sib}.

\end{itemize}

\subsection{Active Galactic Nuclei}

The discovery that most nearby galaxies contain
a black hole and the mass of the black hole is strongly correlated
with the mass of the surrounding spheroid of stars \citep{magor} was possibly one of the
most important discoveries in extragalactic astronomy in recent years,
because it implies the formation of the stars and the central black
hole in a galaxy are connected.
Previous submillimetre observations of high-redshift quasars
have detected $\simeq$5-10\% of these objects \cite{priddey,beelen}, 
and the spectral energyy distributions of the detected quasars imply
that
most of the submillimetre emission must be coming
come from a starburst surrounding the active nucleus rather than directly
from the nucleus itself. This supports
the idea that formation of the stars
and the formation of the central black hole in a galaxy are connected.

We will use the H-ATLAS to detect both individual active galactic nuclei that fall in our fields
and to carry out statistical analyses of AGN that are too faint to detect
individually by coadding (`stacking') the Herschel emission at the positions
of all the AGN in a given class. Let us consider as an example Type 1 quasars.
Figure 11 shows quasars drawn from the Palomar Green (PG) survey and the SDSS
as a function of redshift and optical luminosity. Because of the strong correlation between
luminosity and redshift in any flux-limited sample, it is crucial to observe more than
one sample, so that one can compare quasars with the same optical luminosity at different
redshifts and vice versa. The detection rate of 
PG quasars by IRAS and ISO is $\simeq$80\% \citep{hauss}.
We will use the H-ATLAS to extend the
far-IR/submillimetre study of quasars to the more distant SDSS quasars, of
which there are $\simeq10^4$ in the H-ATLAS fields.
Using the results of a pilot study with the Spitzer legacy survey SWIRE \cite{serjeant}, 
we estimate
that we will detect $\simeq$440 quasars at $z < 3$ and $\simeq 210$ quasars at $z > 3$. 
This is $\simeq$15 times greater than the number of existing detections of high-redshift quasars.
We will investigate the properties of the quasars that are too faint to detect individually
by a stacking analysis.

\begin{figure}
\figurenum{11}
\plotone{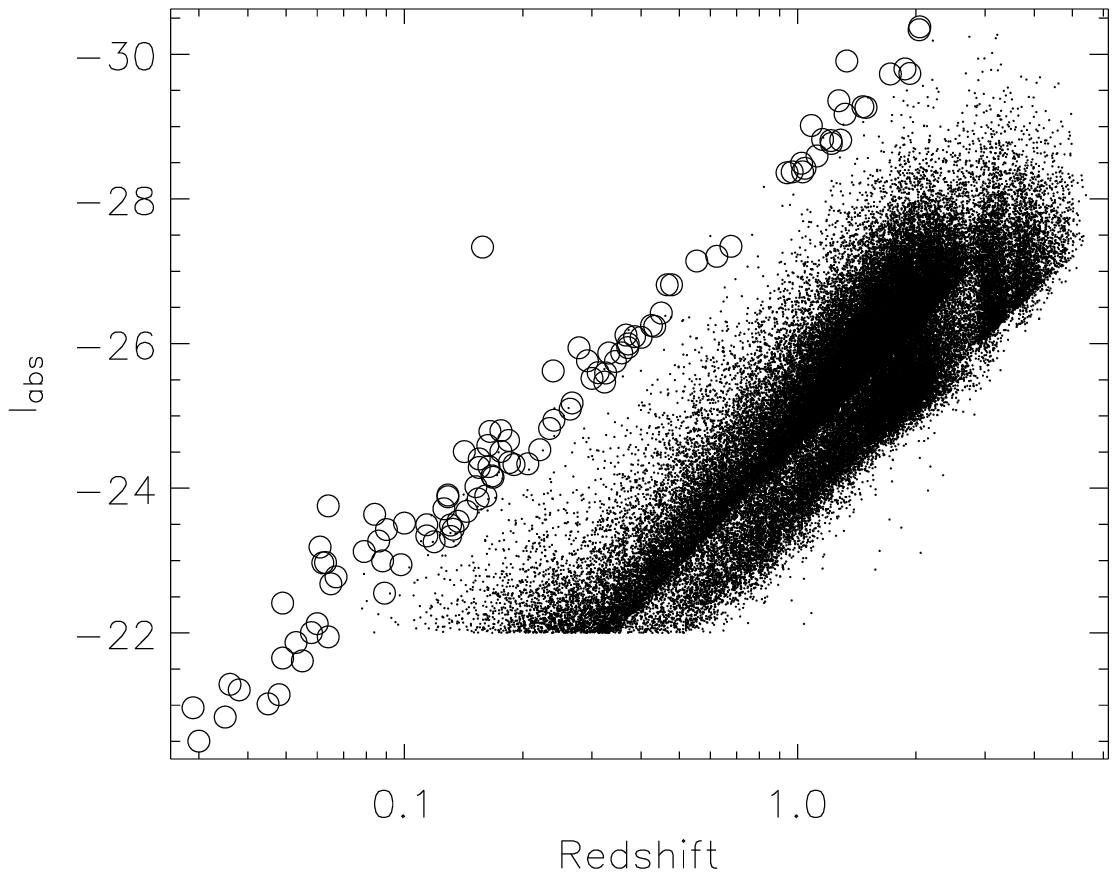}
\caption{The distribution of PG (circle) and SDSS (point)
QSOs as a function of
redshift and luminosity.}
\end{figure}

\subsection{Large-Scale Structure}

The H-ATLAS  will detect $\sim$200,000 sources with a median redshift of $\sim$1 and will
therefore
contain a large amount of information about large-scale structure upto a scale of
$\simeq$1000 Mpc at $\rm z \sim 1$.
Many large-scale structure projects, such as searches for baryonic oscillations, will
only become possible in the future, as the
near-IR (VIKING,UKIDSS) and optical (KIDS, Pan-STARRS, the Dark Energy Survey) surveys
eventually provide photometric redshift estimates for
most of the sources. There are two projects that are possible immediately:

\begin{itemize} 

\item The H-ATLAS will make possible measurements of the angular correlation function
on very large scales. These measurements will make it possible to discriminate
between some
models of galaxy formation. Figure 12, for example, shows the results of using a Monte-Carlo simulation
to predict the amplitude of the
angular correlation function of H-ATLAS sources using three alternative models
of galaxy formation: the hybrid model from Van Kampen et al. (2005) and two new models (Van Kampen et al.
in preparation). The simulation shows that it should be easy to discriminate between these models.

\item The individual sources, however, only represent $\simeq$10\% of the
extragalactic background radiation at the Herschel wavelengths and the unresolved background
contains a wealth of further information about the clustering
properties of dusty galaxies. 
A powerful technique to extract further information is
to investigate the clustering properties of 
the intensity distribution on the maps after the high signal-to-noise
sources have been removed \citep{amblard,lagache2007}.
The strength of the fluctuations on large angular
scales (linear regime) can be used to estimate the average mass of the
dark matter halos containing the sources, whereas the 
strength of the fluctuations on small angular scales (non-linear regime)
can be used to estimate the
halo occupancy distribution \citep{amblard}.

\end{itemize}

\begin{figure}
\figurenum{12}
\plotone{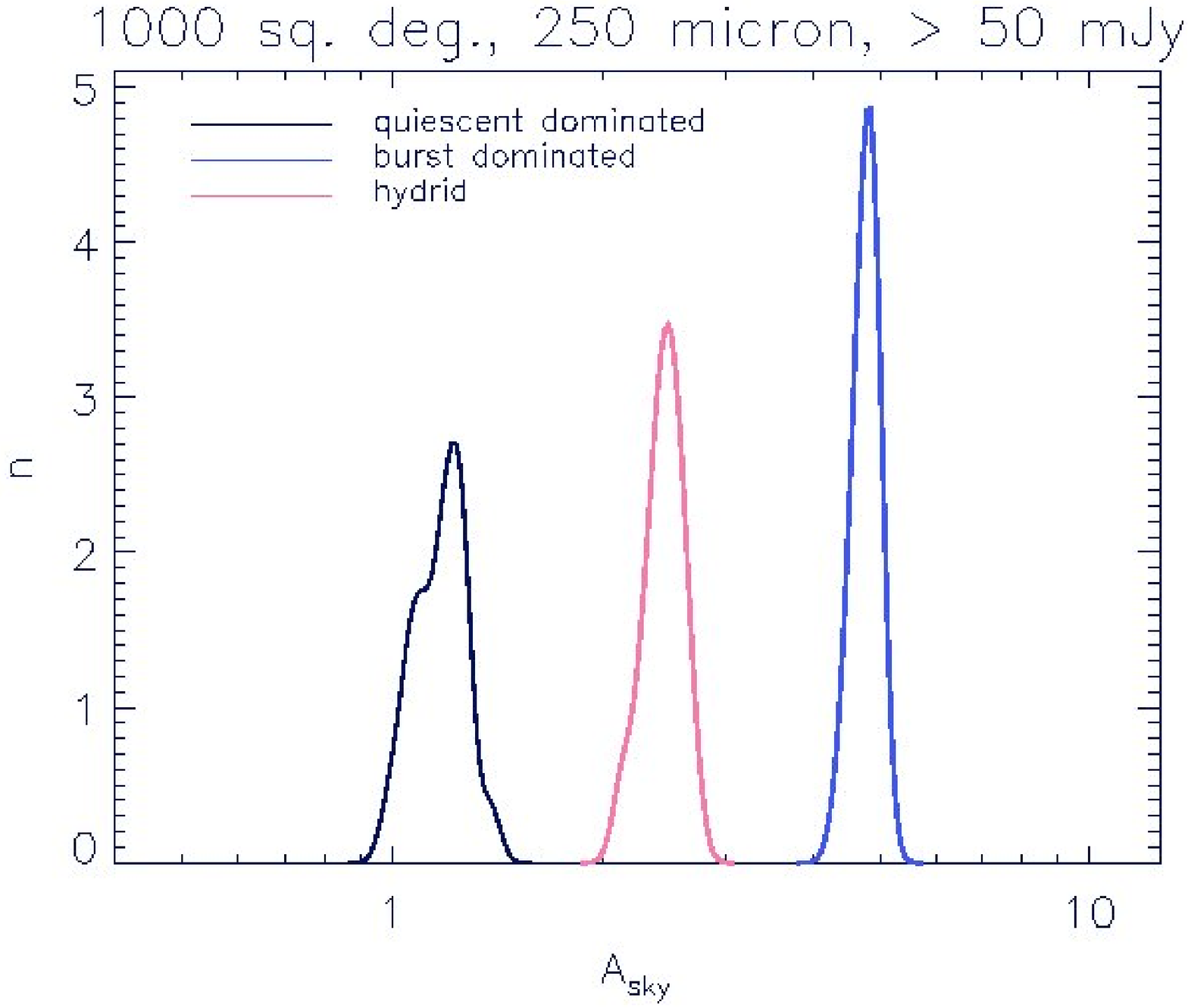}
\caption{A histogram of 
the amplitude of the
angular correlation function ($A_{sky}$) of H-ATLAS sources for
100 Monte-Carlo realisations for each of three different models of
galaxy formation: the hybrid model 
described by Van Kampen et al. (2005) and two new models (Van Kampen et al. in preparation).}
\end{figure}

\subsection{Galactic Dust Sources}

We will use the H-ATLAS to look for dust associated with stars, especially debris disks and dust
around stars on the asymptotic giant branch. We will also study the
distribution and spectral properties of dust in the interstellar medium.
Although the 
H-ATLAS fields are at high galactic latitudes, in order to minimize the emission from dust within
the Galaxy, IRAS images show that there
is plenty of dust even at the galactic poles (Figure 13). The H-ATLAS will 
allow us to study the
structure of an interstellar dust on an angular scale $\simeq$10 times smaller
than was possible
with IRAS \cite{mill} and will be possible with Planck, while still having as good sensitivity to 
large-scale structure as the two other telescopes.
The H-ATLAS will also detect
the cold dust missed by IRAS.

\begin{figure}
\figurenum{13}
\plotone{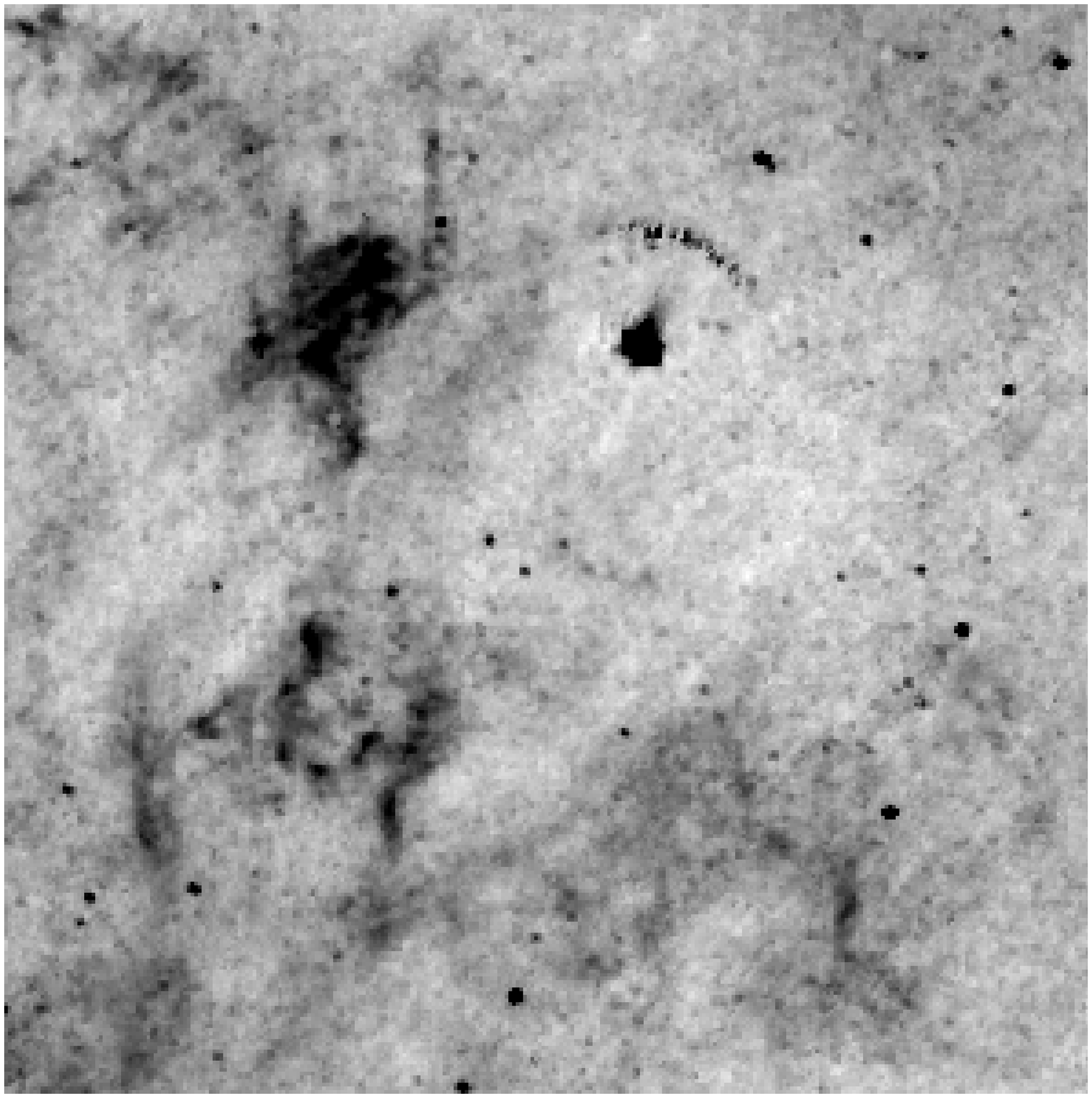}
\caption{IRAS image at 100 $\mu$m of a $\rm 10 \times 10\ deg^2$ region around
the south galactic pole.}
\end{figure}
 
A more speculative idea is to look for prestellar cores and protostars
at high latitudes. Since the dust disks in galaxies are thin, most of the high-latitude
dust is within 0.5 kpc. Therefore, we will be able to carry out a search for prestellar
cores and protostars in our fields down to surprisingly low mass limits. On the assumption
of a dust temperature of 20 K and a standard gas-to-dust ratio, we estimate that we
should detect all protostars and prestellar 
cores down to a mass of $\sim0.002\ M_{\odot}$---well below the 
brown dwarf limit and in the Jupiter regime. 

\section{Further Technical Details and Data Release}

The basic parameters of the survey were given in \S 2.
In this section, we describe some technical issues that will affect
the nature of the data that we eventually release to
the community.

An important feature of the survey is that we plan to observe all fields twice.
The main purpose of this is to overcome the potential problem of low-frequency
(1/f) noise, caused by either slow drifts in the temperatures of the detectors
or other changes in the instrument electronics. These drifts in time can lead to spurious large-scale
structure in the final maps, which would be a problem for some of our science projects.
As long as the two sets of observations are in different directions, however, it should
be possible to use mapping algorithms such as MADmap \cite{cantaloupe}
to produce maximum-likelihood images of sky that are free of the effects of this
noise.
Simulations suggest that as long as the
scan direction of the two sets of observations is separated by at least 20 degrees, 
there should be no features
in the final maps caused by these
drifts \cite{tim2}.

Achieving two sets of observations with different scan directions, however, is quite tricky,
partly because of the severe constraints on the way that Herschel will observe the sky and
partly because of the sheer size of H-ATLAS (11\% of the time available for open-time
key projects).
Surveys will be carried out with Herschel by using the telescope to scan across the sky
along a great circle, then moving the telescope a short distance in a parallel `cross-leg'
direction, and then moving the telescope back along a great circle, thus
gradually building up a map of the sky. A complication of observing with SPIRE
is that the SPIRE bolometers do not instantaenously
fully sample
the sky, and so to produce a fully-sampled SPIRE image it is necessary to carry out the
scan in one of 24 possible directions that are related to the six-fold
symmetry of the array (there is not a similar constraint for
PACS because it is a fully-sampled array). Other complications are that it is necessary to keep the
Sun safely behind the telescope's sun-shield and that it is only possible to rotate the telescope completely
around one axis.
Waskett et al. (2007) give a good introduction to the
complications of carrying out surveys with Herschel. The result of these constraints
is that it is only possible to observe a given field during certain visibility windows, the duration of which
depends on the position of the field. 

The easiest fields to observe are the GAMA fields, which have long visibility windows because they
are the ones closest to the ecliptic plane. The
Astronomical Observation Requests (AOR), the units out of which the telescope's observing programme is built,
of the GAMA fields each consist of two sets of observations with orthogonal scan directions of an area of sky $\simeq \rm 4 \times 4\ deg^2$. 
These AORs, which will take approximately 16 hours to complete, will 
result in approximately square images, like those shown in Figure 2, although the orientation of the
square will depend on the exact time at which the observation is made. Thus the coverage within each square will be
uniform but we cannot predict at the moment the angle that each square will make to the celestial equator, and 
thus the precise overlap with the area covered by the surveys in other wavebands (\S 4).

\begin{figure}
\figurenum{14}
\plotone{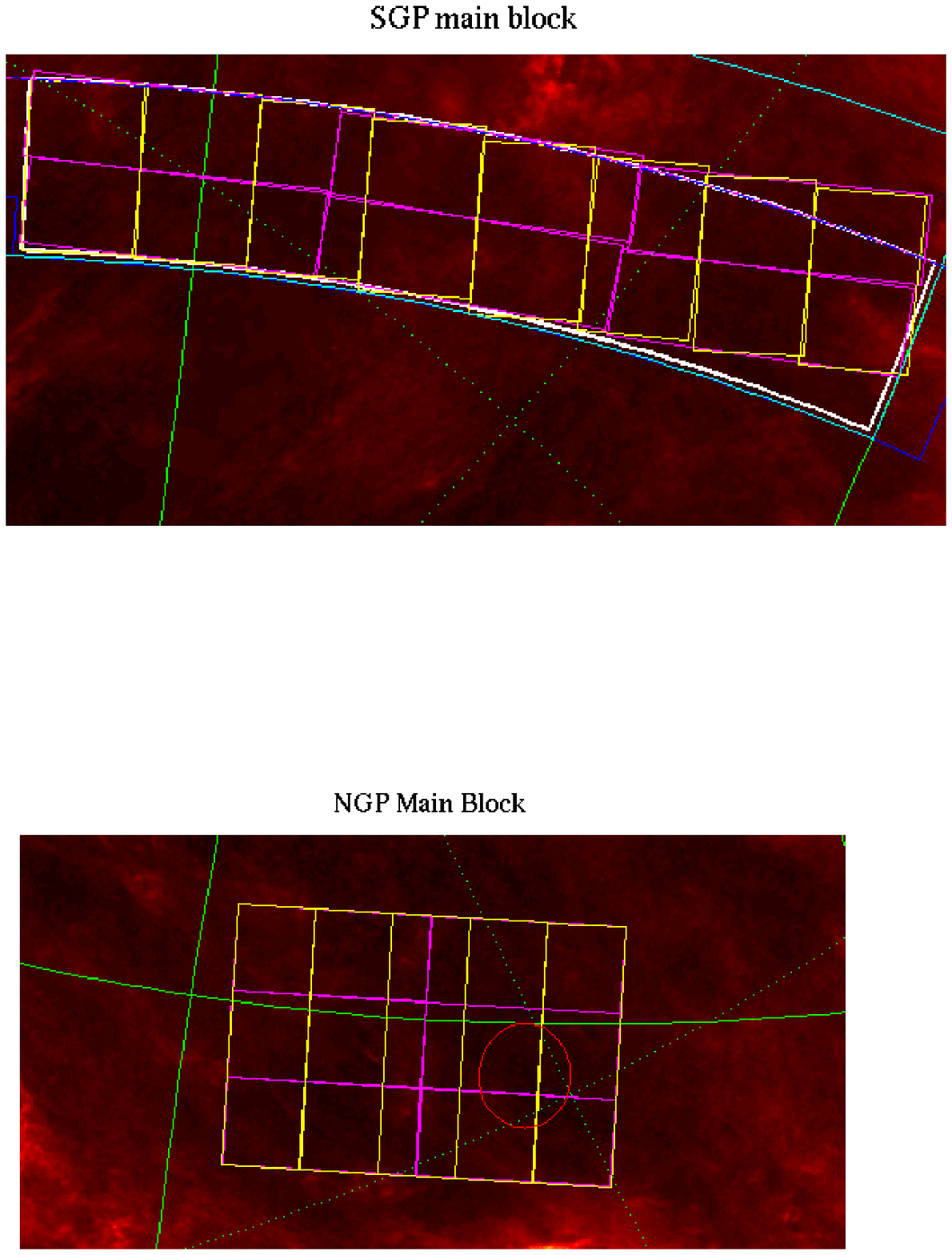}
\caption{The design of the surveys of the ATLAS NGP field and of the larger of the two
SGP fields. In both cases, the purple and yellow lines show the areas that will be covered
by individual Astronomical Observation Requests (AORs). The other colours outline the areas
covered by complementary datasets in other wavebands. The red circle in the NGP figure
shows the position of the Coma Cluster.}
\end{figure}

The other two fields are more difficult to observe because they have shorter visibility windows and because we need
to map large contiguous areas of sky to make the best use of the complementary data in other
wavebands. As an example, we will consider the larger of the fields near the south galactic pole (SGP).
To make the best use of the redshifts from the 2dF Galaxy Redshift Survey, the field has 
a small range of declination but a large range of right ascension (Fig. 2). For this field, the
individual AORS each contain one set of observations with a single scan direction.
Figure 14 shows the layout of the AORs,
although their exact orientation depends on exactly when the
observations are done.
The scan directions are either roughly along lines of constant declination, with the observations covering
a rectangular area of $\rm \simeq3^{\circ}$ in declination and $\rm \simeq 12^{\circ}$ in right ascension, or
roughly along lines of constant right ascension, with the observations covering a rectangular area of
$\rm \simeq 6^{\circ}$ degrees in declination and $\rm \simeq 4^{\circ}$ in right ascension. 
We have set some constraints on when the AORs are observed to avoid
the rectangles being rotated too far from a north-south line but it is not possible to make the constraints too
severe because of the difficulty of scheduling such a large programme. Fig 15 shows an example of the coverage
we are likely to have for this field. The full range of the exposure time shown by the colour table is a factor of four,
so the sensitivity in the apparent gaps is 
a factor of $\rm \simeq 2$ worse than in the places where we have the best coverage.
Apart from in these gaps, every point in the field will have at least two sets 
of observations, with the scans in approximately
orthogonal directions, and thus we will be able to use maximum-likelihood imaging algorithms to remove the
effect of $1/f$ noise.
Figure 14 also shows the design of the AORs for the field near the north galactic pole (NGP). 
The individual AORS again each contain one set of observations with a single scan direction.
There are two sets of AORs with roughly orthogonal scan directions, and each covering a rectangular
area of $\rm \simeq 9^{\circ} \times 4^{\circ}$ with the longer side in the scan direction.
The orientation of the overall field, and the coverage within the field, is still uncertain and will
depend on the final telescope schedule. For all the fields, one of the legacy data products will be coverage maps of the
fields.

\begin{figure}
\figurenum{15}
\plotone{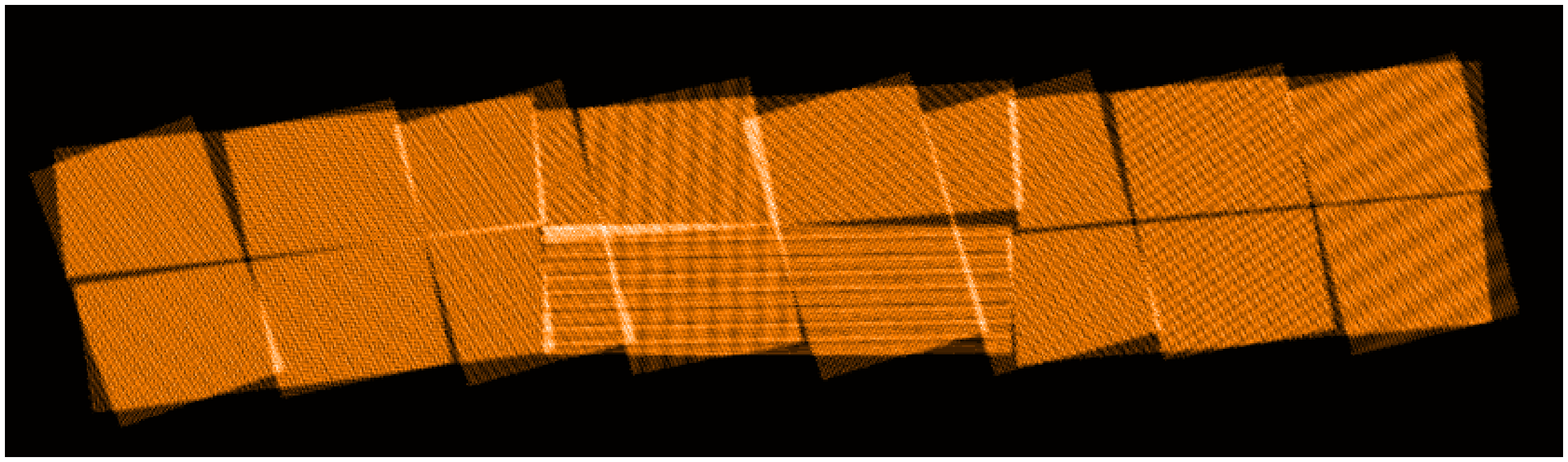}
\caption{A simulation of the likely coverage of the larger of the fields near the south galactic pole.
The range of exposure time shown by the colour table is a factor of four.}
\end{figure}

A useful byproduct of our observing strategy of two sets of observations of each point with an
field is that we will be able to look for time-varying and moving objects in the H-ATLAS data.
Because of the scheduling difficulties, we have not placed 
any constraint on the time interval between two observations of the same place within the SGP
and NGP fields, and thus it is only within the GAMA fields that we will have two sets of observations separated
in time by a roughly constant interval. This interval is approximately 8 hours, which is well-suited
for looking for asteroids but not, for example, for objects in the Edgeworth-Kuiper Belt.

We intend to release the H-ATLAS data (maps and catalogues) to the
community in a series of data releases.
The first data release will be for a $\rm 4^{\circ} \times 4^{\circ}$ area within the
9-hour GAMA field (Table 1), which will be
observed during the Herschel Science Demonstration Phase (October 15th-November 30th 2009).
We currently expect to release the maps and catalogues for this region in May 2010.
The
complexity of telescope scheduling means that we cannot yet give a firm date
for the next data release. More information can be found on the Herschel ATLAS website (www.h-atlas.org).

\section{Conclusions}

The Herschel ATLAS is the largest open-time key project that will be carried out with the
Herschel Space Observatory. It will survey 510 square degrees of the extragalactic
sky, four times larger than all the other Herschel surveys combined, in five far-infrared
and submillimetre bands.
We have described the survey, the complementary multi-wavelength datasets that will
be combined with the Herschel data, and the six major science programmes we plan
to undertake.
Using new models based on a submillimetre survey of nearby galaxies, we have presented predictions of the properties
of the sources
that will be detected by the H-ATLAS. We intend to release the H-ATLAS data---maps and
catalogues---to the astronomical community in a series of data releases. The first
of these will be in May 2010.

\acknowledgements

We thank the UK Science and Technology Facilities Council and the Italian Space Agency (ASI contract
I/016/07/0 COFIS) for funding.
This research has made use of the NASA/IPAC Extragalactic Database (NED), which is operated by
the Jet Propulsion Laboratory, California Institute of Technology, under contract with the
National Aeronautics and Space Administration.





\clearpage

\end{document}